\documentclass[useAMS,usenatbib]{mn2e}
\usepackage{times}

\newcommand{\um}{\,$\mu$m}

\newcommand{\td}{$T_{\rm{d}}$}
\newcommand{\msun}{\,$\rm{M}_{\odot}$}
\newcommand{\lsun}{\,$\rm{L}_{\odot}$}
\def\gs{\mathrel{\raise0.35ex\hbox{$\scriptstyle >$}\kern-0.6em\lower0.40ex\hbox{{$\scriptstyle \sim$}}}}
\def\ls{\mathrel{\raise0.35ex\hbox{$\scriptstyle <$}\kern-0.6em\lower0.40ex\hbox{{$\scriptstyle \sim$}}}}

\title{The 1\,--\,1000\um\ SEDs of far-infrared galaxies}
\author[Sajina et al.] {A.~Sajina$^{1,2}$, D.~Scott$^2$,  M.~Dennefeld$^3$, H.~Dole$^{4,5}$, M.~Lacy$^1$, G.~Lagache$^4$ \\
  $^1${\sl Spitzer} Science Center, California Institute of Technology, Pasadena, CA 91125, USA\\
  $^2$Department of Physics \& Astronomy, University of British Columbia, 6224 Agricultural Road, Vancouver, BC V6T1Z1, Canada\\
  $^3$Institut d'Astrophysique de Paris, 98bis Boulevard Arago, Paris, France\\
  $^4$Institut d'Astrophysique Spatiale, Universite Paris Sud, Bat. 121,  91405 Orsay Cedex, France\\
  $^5$Steward Observatory, University of Arizona, 933 N Cherry Ave, Tucson, AZ 85721, USA\\  	}

\pagerange{\pageref{firstpage}--\pageref{lastpage}}
\pubyear{0000}

\begin{document}

\pagerange{\pageref{firstpage}--\pageref{lastpage}} \pubyear{2006}

\maketitle

\label{firstpage}

\begin{abstract}
Galaxies selected at 170\um\ by the {\sl ISO} FIRBACK survey represent the brightest $\sim$\,10\% of the Cosmic Infrared Background. Examining their nature in detail is therefore crucial for constraining models of galaxy evolution. Here we combine {\sl Spitzer} archival data with previous near-IR, far-IR, and sub-mm observations of a representative sample of 22 FIRBACK galaxies spanning three orders of magnitude in infrared luminosity. We fit a flexible, multi-component, empirical SED model of star-forming galaxies designed to model the entire $\sim$\,1\,--\,1000\um\ wavelength range. The fits are performed with a Markov Chain Monte Carlo (MCMC) approach, allowing for meaningful uncertainties to be derived. This approach also highlights degeneracies such as between \td\ and $\beta$, which we discuss in detail. From these fits and standard relations we derive: $L_{\rm{IR}}$, $L_{\rm{PAH}}$, SFR, $\tau_{\rm{V}}$, $M_{*}$, $M_{\rm{dust}}$, \td, and $\beta$. We look at a variety of correlations between these and combinations thereof in order to examine the physical nature of these galaxies. Our conclusions are supplemented by morphological examination of the sources, and comparison with local samples. We find the bulk of our sample to be consistent with fairly standard size and mass disk galaxies with somewhat enhanced star-formation relative to local spirals, but likely not bona fide starbursts. A few higher-$z$ LIGs and ULIGs are also present, but contrary to expectation, they are weak mid-IR emitters and overall are consistent with star-formation over an extended cold region rather than concentrated in the nuclear regions.    
We discuss the implications of this study for understanding populations detected at other wavelengths, such as the bright 850\um\ SCUBA sources or the faint {\sl Spitzer} 24\um\ sources. 
\end{abstract}


\section{Introduction}
Understanding the full infrared spectral energy distribution (SED) of galaxies is essential for a complete picture of star-formation in the Universe. Measurements of the cosmic background radiation allow us to infer that about half the energy ever generated by stars was reprocessed by dust into the infrared \citep[see][for a review]{hd01}. This emission is increasingly important at higher redshifts where the star-formation density of the Universe is larger than today. Modelling the contribution of different galaxy populations to the Cosmic Infrared Background (CIB) requires detailed knowledge of the SEDs of star-forming galaxies. The variation in SED shapes is a key uncertainty when comparing populations selected at different wavelengths or testing galaxy evolution models. \\
In this paper, we discuss the full infrared SED ($\sim$\,1\,--\,1000\um), which roughly spans the wavelengths between stellar and synchrotron-dominated emission. Traditionally, studying this entire range at once has been difficult, since mid-IR ($<$\,60\um) observations could not reach much beyond the local Universe\footnote{Except for the deepest ISOCAM 15\um\ observations which peak at $z$\,$\sim$\,0.7 \citep{rod04}}, while sub-mm observations also only exist for very local, IR-bright, galaxies \citep[e.g.][]{de01}, or else the blank-sky Sub-mm Common User Bolometer Array (SCUBA) sources which peak at $z$\,$\sim$\,2--3 \citep{c05}. The latter typically have only one or two other wavelength detections in addition to the SCUBA (850\um) one, which makes their interpretation particularly dependent on the SED model assumed \citep[see][for a review]{bl02}.  Due to these past observational limitations, we still do not know (beyond some generalized trends) the full range of galaxy SED shapes, how exactly they are related to the underlying physical conditions in the galaxy, and therefore how they may vary across cosmic time as the galaxies evolve.  \\ 
With the advent of the {\sl Spitzer} Space Telescope \citep{wer04} we can for the first time observe the mid-IR properties of large numbers of sources over a cosmologically significant extent in redshift (e.g.~Lonsdale et al.~2003). {\sl Spitzer} covers the range 3--160\um. The obvious next step toward characterizing the full infrared SEDs of galaxies is therefore to link {\sl Spitzer} observations with longer wavelength samples, especially including sub-mm observations. In this regard, the quality and quantity of the available sub-mm data are the limiting factor. \\
The sample discussed here is selected from the 170\um\ FIRBACK (Far-IR BACKground) ELAIS-N1 catalog \citep{p99,d01}. The selection is based on an existing radio detection, which we followed-up with both deep near-IR and sub-mm observations (Scott et al.~2000; Sajina et al.~2003, hereafter S03). We find that in terms of the mid-IR properties, the sample selection is largely unbiased with respect to the FIRBACK population as a whole. Our previous studies suggest that the sample consists primarily of $z$\,$<$\,0.3 ordinary, spiral-like galaxies, rich in cold ($T$\,$<$\,40\,K) dust, with $\sim$1/6 of the sample consisting of Ultraluminous Infrared Galaxies (ULIGs) at $z$\,$\sim$\,0.5\,--\,1. Spectroscopic follow-up of this and related sub-samples support these findings \citep[hereafter D05]{c02,patris,den05}. The higher-$z$ fraction therefore represents a bridge population between the local Universe and distant, dusty star-formers such as the SCUBA blank-sky sources. In general, the importance of FIRBACK galaxies is that they represent the brightest galaxies at 170\um, contributing $\sim$\,10\% to the CIB and {\it selected at a wavelength near the peak of the CIB spectrum}. \\
We use archival {\sl Spitzer} observations of the ELAIS-N1 field in order to extend the known SEDs of the above sample into the mid-IR wavelength range. We fit these SEDs with a phenomenological model motivated by different physical origins for the emission. These fits allow us to discuss the physical characteristics of our galaxies as well as trends within the sample. Details of the fitting procedure and some related issues are included in a set of appendices. \\ 
Throughout the paper we assume a flat Universe with $H_{0}$\,=\,75~km~s$^{-1}$~Mpc$^{-1}$, $\Omega_{\rm{M}}$\,=\,0.3, and $\Omega_{\Lambda}$\,=\,0.7. 
\begin{figure}
\centering
\vspace*{6.5cm}
\leavevmode
\includegraphics{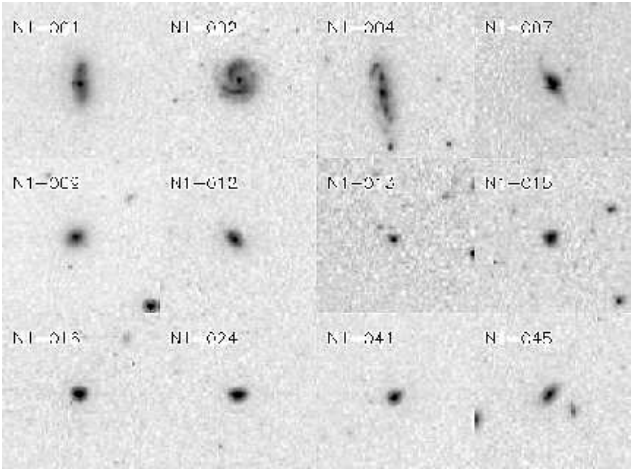}
\caption{\label{mosaic}  IRAC 3.6\um\ images for bright galaxies in our sample. Note the prevalence of disk-like morphologies with bright nuclei (N1-004 is a spectroscopically confirmed weak AGN). Boxes here are approximately 1.4\,arcmin wide. For comparison, {\sl ISO}'s 170\um\ beam is $\sim$\,90\,arcsec. }
\end{figure} 

\section{Data}
\label{secdata}
The galaxies we focus on come from a sub-sample of FIRBACK sources with radio detections which were observed with SCUBA. The full sample of 30 targets was first presented in S03 where details of the sample selection as well as the near-IR ({\em K}) and sub-mm observations and data reduction are given. In particular, in S03 we addressed the question of whether or not the radio detection requirement introduces a bias and found that apart from potential HyLIGs ($L_{\rm{IR}}$\,$>$\,$10^{13}$\lsun) at $z$\,$\sim$\,1.5, and spurrious sources, the radio detection does not bias us with respect to the FIRBACK catalogue as a whole.  Here we additionally use {\sl Spitzer} archival observations of the ELAIS-N1 field obtained as part of the {\sl Spitzer} Wide-area IR Extragalactic Survey (SWIRE, Lonsdale et al.~2003). We focus on sources with unique identifications in the SWIRE data (see Section~2.2), which leaves us with a sample of 22 sources (see Table~\ref{mwdata}). 
\subsection{Treatment of {\sl Spitzer} data}
The basic data reduction and calibration is already performed on the individual frames obtained from the {\sl Spitzer} archive. We use the Starlink software package {\sc ccdpack} to rescale, align and coadd the observations into common mosaic images for each of the 4 IRAC bands and the MIPS 24\um\ band. Furthermore, the Starlink {\sc photom} and {\sc gaia} packages are used to perform aperture photometry on the sources. To ensure that the sky-annuli chosen fairly represent the sky/background we monitor the sky values obtained for all of these and modify the annular region for any galaxy whose sky value is more than 2\,$\sigma$ above the average sky. The errors are then computed from the sky variance. We find unambiguous counterparts for nearly all radio/SCUBA sources in S03 (the exceptions being N1-032, and N1-034). \\ 
\begin{table*}
\centering
\begin{minipage}{180mm}
\caption{\label{mwdata} Multiwavelength data for our sample. All errors are 1\,$\sigma$ estimates.}
\scriptsize
\begin{tabular}{lrrrrrrrrrrrr}
Source$^*$ & $S_{1.3}$$^a$ & $S_{2.2}$$^b$ & $S_{3.6}$$^c$ & $S_{4.5}$$^c$ & $S_{5.8}$$^c$ & $S_{8.0}$$^c$ & $S_{24}$$^c$ & $S_{70}$$^c$ & $S_{170}$$^d$ & $S_{450}$$^b$ & $S_{850}$$^b$ & $S_{1.4\rm{GHz}}$$^e$  \\
 & mJy & mJy & mJy & mJy & mJy & mJy & mJy & mJy & mJy & mJy & mJy & mJy \\
 & & & & &  & & & & & & \\
N1-001 & 6.28 & 7.31$\pm$0.54 & 2.97$\pm$0.32 & 1.98$\pm$0.26 & 4.67$\pm$0.4 & 12.25$\pm$0.66 & 19.97$\pm$1.71 & 233 & 597$\pm$72 & --3.0$\pm$14.0 & 6.1$\pm$1.6 & 0.74$\pm$0.23 \\
N1-002 & 4.58 & 5.55$\pm$0.41 & 5.35$\pm$0.43 & 3.79$\pm$0.37 & 9.6$\pm$0.59 & 23.97$\pm$0.92 & 9.87$\pm$1.21 & 112 & 544$\pm$69 & 14.4$\pm$12.4 & 4.4$\pm$1.1 & 0.64$\pm$0.04 \\
N1-004 & 9.16 & 7.31$\pm$0.14 & 5.38$\pm$0.44 & 3.79$\pm$0.36 & 6.05$\pm$0.47 & 11.42$\pm$0.64 & --- & --- & 391$\pm$58 & 32.5$\pm$7.2 & 3.6$\pm$1.4 & 0.88$\pm$0.13 \\
N1-007 & 4.40 & 3.5$\pm$0.26 & 3.65$\pm$0.36 & 2.41$\pm$0.29 & 4.09$\pm$0.38 & 11.87$\pm$0.65 & --- &  --- & 338$\pm$54 & 23.4$\pm$8.1 & 4.4$\pm$1.6 & 1.04$\pm$0.12 \\
N1-009 & 9.63 & 10.57$\pm$0.2 & 6.02$\pm$0.46 & 3.96$\pm$0.37 & 7.05$\pm$0.5 & 14.03$\pm$0.7 & 4.26$\pm$0.8 & 96 & 313$\pm$52 & 10.6$\pm$7.6 & 3.5$\pm$1.5 & 1.15$\pm$0.11 \\
N1-012 & 1.75 & 1.84$\pm$0.3 & 1.15$\pm$0.2 & 0.75$\pm$0.16 & 0.81$\pm$0.17 & 4.24$\pm$0.39 & --- &  --- & 302$\pm$51 & 9.2$\pm$10.0 & 1.5$\pm$1.6 & 0.31$\pm$0.07 \\
N1-013 & --- & 0.13$\pm$0.01 & 0.16$\pm$0.07 & 0.14$\pm$0.07 & 0.18$\pm$0.08 & 0.44$\pm$0.12 & 2.21$\pm$0.58 & 83 & 294$\pm$51 & 18.8$\pm$9.9 & 0.0$\pm$1.5 & 0.52$\pm$0.15 \\
N1-015 & 0.36 & 0.80$\pm$0.06 & 0.57$\pm$0.14 & 0.53$\pm$0.14 & 0.37$\pm$0.11 & 2.05$\pm$0.27 & 3.16$\pm$0.68 & 49 & 294$\pm$51 & --3.4$\pm$7.7 & 1.4$\pm$1.6 & 0.52$\pm$0.07 \\
N1-016 & 2.71 & 3.5$\pm$0.26 & 1.78$\pm$0.25 & 1.41$\pm$0.22 & 1.61$\pm$0.24 & 7.6$\pm$0.52 & 6.81$\pm$1 & 160 & 289$\pm$50 & 34.8$\pm$16.7 & 1.5$\pm$1.2 & 1.55$\pm$0.13 \\
N1-024 & 1.14 & 1.39$\pm$0.03 & 1.16$\pm$0.2 & 0.77$\pm$0.16 & 0.9$\pm$0.18 & 5.4$\pm$0.44 & 4.38$\pm$0.81 & 108 & 266$\pm$49 & 32.3$\pm$7.5 & 2.9$\pm$1.3 & 0.75$\pm$0.02 \\
N1-029 & 1.10 & 1.27$\pm$0.09 & 1.24$\pm$0.21 & 0.88$\pm$0.18 & 0.75$\pm$0.16 & 4.25$\pm$0.39 & 4.3$\pm$0.8 & 72 & 229$\pm$46 & 20.0$\pm$14.2 & 0.5$\pm$1.7 & 0.69$\pm$0.05 \\
N1-031 & 1.93 & 2.65$\pm$0.19 & 1.45$\pm$0.22 & 0.91$\pm$0.18 & 1.44$\pm$0.23 & 3.53$\pm$0.35 & 5.41$\pm$0.9 & 62 & 225$\pm$46 & 9.2$\pm$13.2 & 1.9$\pm$1.1 & 0.43$\pm$0.06 \\
N1-039 & --- & 0.32$\pm$0.05 & 0.28$\pm$0.1 & 0.25$\pm$0.09 & 0.19$\pm$0.08 & 0.8$\pm$0.17 & 1.3$\pm$0.44 & 44 & 205$\pm$44 & 10.9$\pm$86.8 & --0.1$\pm$2.3 & 0.58$\pm$0.07 \\
N1-040 & --- & 0.01$\pm$0.01 & 0.08$\pm$0.05 & 0.07$\pm$0.05 & 0.06$\pm$0.05 & 0.05$\pm$0.04 & 0.69$\pm$0.32 &  26 & 205$\pm$44 & 29.2$\pm$20.5 & 5.4$\pm$1.1 & 0.33$\pm$0.03 \\
N1-041 & 0.38 & 0.88$\pm$0.06 & 0.66$\pm$0.15 & 0.5$\pm$0.13 & 0.58$\pm$0.14 & 4.14$\pm$0.38 & 3.89$\pm$0.76 & 75 & 204$\pm$44 & 20.4$\pm$156.7 & --0.1$\pm$2.5 & 0.76$\pm$0.06 \\
N1-045 & 1.39 & 1.16$\pm$0.02 & 1.01$\pm$0.19 & 0.66$\pm$0.15 & 0.69$\pm$0.16 & 3.29$\pm$0.34 & 2.23$\pm$0.58 & 64 & 198$\pm$44 & 15.3$\pm$8.3 & 3.0$\pm$1.4 & 0.43$\pm$0.06 \\
N1-064 & --- & 0.03$\pm$0.01 & 0.15$\pm$0.07 & 0.11$\pm$0.06 & 0.12$\pm$0.07 & 0.09$\pm$0.06 & 1.38$\pm$0.46 &  14 & 166$\pm$42 & 35.2$\pm$13.9 & 5.1$\pm$1.2 & 0.23$\pm$0.04 \\
N1-068 & 0.31 & 0.51$\pm$0.04 & 0.31$\pm$0.1 & 0.31$\pm$0.11 & 0.28$\pm$0.1 & 1.98$\pm$0.26 & 3.96$\pm$0.76 & 62 & 165$\pm$42 & 15.1$\pm$7.6 & 2.2$\pm$1.4 & 0.44$\pm$0.05 \\
N1-077 & 0.31 & 0.42$\pm$0.07 & 0.37$\pm$0.11 & 0.29$\pm$0.1 & 0.23$\pm$0.09 & 1.4$\pm$0.22 & 1.29$\pm$0.44 &  42 & 159$\pm$41 & 5.9$\pm$7.3 & 1.1$\pm$1.3 & 0.40$\pm$0.10 \\
N1-078 & --- & 0.04$\pm$0.01 & 0.13$\pm$0.07 & 0.1$\pm$0.06 & 0.09$\pm$0.06 & 0.13$\pm$0.07 & 0.6$\pm$0.3 &  --- & 158$\pm$41 & 35.2$\pm$8.7 & 5.7$\pm$1.3 & 0.24$\pm$0.04 \\
N1-083 & --- & 0.46$\pm$0.08 & 0.31$\pm$0.1 & 0.25$\pm$0.09 & 0.18$\pm$0.08 & 0.68$\pm$0.16 & 1.21$\pm$0.43 & 43 & 150$\pm$41 & 16.2$\pm$16.0 & 0.7$\pm$1.2 & 0.55$\pm$0.03 \\
N1-101 & 0.38 & 0.55$\pm$0.09 & 0.58$\pm$0.14 & 0.38$\pm$0.12 & 0.46$\pm$0.13 & 2.34$\pm$0.29 & 2.66$\pm$0.63 & 46 & 136$\pm$40 & 19.8$\pm$7.5 & 0.9$\pm$1.5 & 0.39$\pm$0.05 \\
 & & & & &  &  & & & & & &\\
\end{tabular}
\medskip

$^*$ The naming scheme follows Dole et al.~(2001).\\
$^a$ $\mathrel{J}$-band magnitudes from Rowan-Robinson et al.~(2004) where 20\% uncertainty is assumed. \\
$^b$ UKIRT and SCUBA fluxes from Sajina et al.~(2003). \\
$^c$ IRAC and MIPS fluxes from archival SWIRE observations, this work; $S_{70}$ from the SWIRE catalogue, all assumed to have an uncertainty of 20\%. \\
$^d$ ISOPHOT 170\,$\mu$m data from Dole et al. (2001). \\
$^e$ VLA 21\,cm fluxes from Ciliegi et al. (1999). \\
\end{minipage}
\end{table*}
Since we started on this project, the SWIRE team has released their own catalogues of the field \citep{swire_cat}. These allow us to double check our IRAC and MIPS 24\um\ photometry, as well as add a 70\um\ point where available. For N1-004, N1-007, and N1-012 there are no 24\um\ or 70\um\ fluxes due to missing data. In addition to the SWIRE 70\um\ catalogue, we extract fluxes for three faint sources: N1-040, N1-064, and N1-077. In all cases, the aperture resulting in the maximum flux was used in order to ensure all emission is accounted for. The only remaining source is N1-078 which is near the edge of the image and thus a confident flux cannot be obtained. \\
This is the only source without any data points between 24\um\ and 170\um. For all of N1-004, N1-007, and N1-012 we have ISOCAM 15\um\ and {\sl IRAS} 60\um\ fluxes compensating for the missing MIPS data. For the few cases which have both a 60\um\ (see D05) and 70\um\ detection (N1-001, N1-002, and N1-016), we find that the 60\um\ fluxes are somewhat higher than the 70\um\ ones contrary to any reasonable SED (given the $S_{70}/S_{170}$ colours); but the difference is within the 20\% calibration uncertainty assigned to the 70\um\ flux. The SWIRE catalogue 70\um\ flux for N1-001 (198\,mJy) was most discrepant originally; however, it is an extended source and we find that a flux of 233\,mJy is more accurate. This 20\% difference is the most severe we expect due to aperture effects for this sample.  We do not explicitly use the few available 160\um\ points, but within the uncertainties they are consistent with the {\sl ISO} 170\um\ points. \\
Near-IR {\em J}-band data for about half of our sample are available from the band-merged ELAIS catalog \citep{mrr04}, while the {\em K}-band data come from our previous work (S03). To convert to flux densities, we used zero points of 1600\,Jy and 670\,Jy for the {\em J}- and {\em K}-bands respectively. \\
The {\sl Spitzer} fluxes are presented in Table~\ref{mwdata} along with the rest of the multiwavelength data used here. A few of the sources have some (or all of) {\em U, G, R}, ISOCAM 15\um, {\sl IRAS} 60\um\ and 100\um\ fluxes, which are given in D05. In our present study, these are used primarily for consistency checks.     

\begin{figure}
\centering
\vspace*{8cm}
\leavevmode
\includegraphics{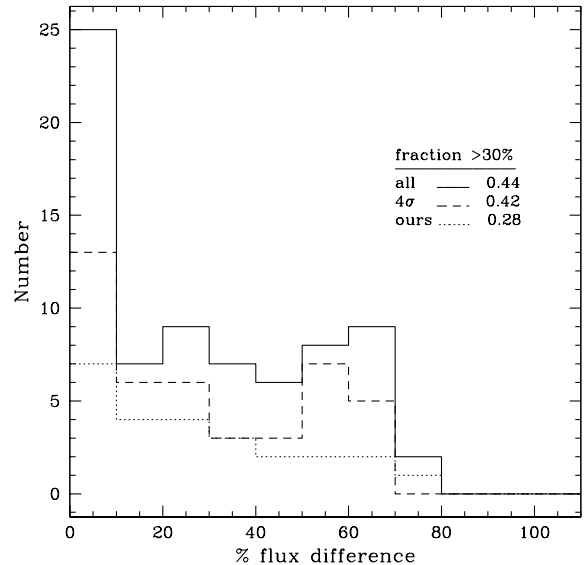}
\caption{\label{diffplot} Quantitative study of the potential effect of multiple sources within the {\sl ISO} beam at 170\um. These histograms show the distribution of flux differences at 24\um\ between using just the brightest 24\um\ source within the {\sl ISO} beam versus using all sources within a 45\,arcsec error circle (i.e. $100$\,$(1-S_{24,max}/\sum S_{24})$). The solid line is for all FIRBACK sources, while the dashed line is for the sources in the $>$\,4\,$\sigma$ catalog. The dotted line is for our sub-sample of 30 radio-selected sources.}
\end{figure}  

\subsection{Possibility of multiple sources}
We use the SWIRE ELAIS-N1 catalog to further investigate the nature of the FIRBACK sources and in particular the relation of our radio-selected sources to the full sample. The 90\,arcsec {\sl ISO} beam means that the chance of multiple sources contributing to the 170\um\ flux is not expected to be negligible. We perform a simple test of this effect by taking every 24\um\ source within a 45\,arcsec radius of the {\sl ISO} position. We then add all these fluxes and compare with the single brightest source within the centroid. The results are presented in Fig.~\ref{diffplot}, where we see that indeed nearly half of all sources $regardless$ $of$ $whether$ $the$ $full$ $or$ $4\sigma$ $catalog$ $is$ $used$ are not well described by assigning a single counterpart. This is a useful qualitative test, although it is unlikely to be correct in detail, because of complications such as foreground AGB stars, or combinations such as a bright local galaxy near a somewhat higher-$z$ LIG/ULIG (where the latter might still contribute the bulk of the 170\um\ flux). Nevertheless, it is evident that nearly half of the 170\um-selected sources have 2 or 3 roughly equivalent mid-IR counterparts. Considering the position of the counterparts within the beam does not appear to affect this significantly. This fraction drops slightly for our radio-selected sample, although it is still significant. This general conclusion is in agreement with the results of Dennefeld et al.~(2005) who find 28/56 sources in the 4\,$\sigma$ catalog to have firm, unique identifications. Here we add that this fraction is unlikely to change significantly for the 3\,$\sigma$ catalog. We also find similar results for our FIRBACK sub-sample of 30 targets, with 28\% having fainter 24\um\ sources in the {\sl ISO} beam contributing $>$\,30\% of the flux of the brightest source (dotted line in Fig.~\ref{diffplot}). \\
Because of strong ambiguities, we remove 8 FIRBACK sources from our sub-sample. We leave 2 ambiguous sources (N1-015 and N1-039) because they have $z_{\rm{spec}}$\,$\sim$\,0.2\,--\,0.3 and this test is inconclusive for higher-$z$ sources. This leaves 22 sources for which the correct counterpart is reasonably secure, with other possible counterparts contributing a probable amount to the 170\um\ flux which is of order the flux uncertainties or less. Since the beamsize at the other wavelengths is so much better than at 170\um, the effects of multiple source contributors to the other fluxes are negligible. The effects of multiple contributors are already partially included in the 170\um\ uncertainties, which contain confusion noise (see Lagache \& Puget~2000). Therefore after removing the 8 {\sl ISO} sources which have the most ambiguous identifications at 24\um, we are confident that the effects of multiple counterparts are not significant. We concentrate on this new sub-sample of 22 sources in the rest of this paper. 
\begin{figure}
\centering
\vspace*{8cm}
\leavevmode
\includegraphics{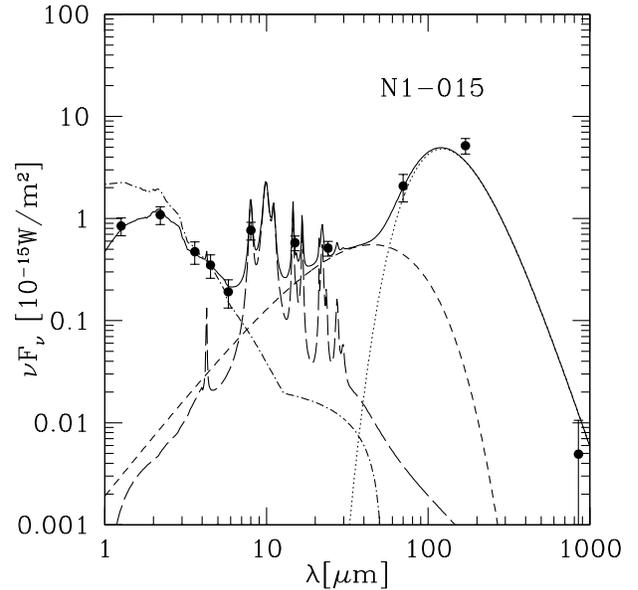}
\caption{\label{sed_demo} An example of the SED model used here. This includes a greybody ({\it dotted line}), a warm power-law ({\it short-dash}), PAH emission ({\it long-dash}), and unextincted stellar emission ({\it dot-dash}) with $e^{-\tau_\nu}$ extinction applied. The thick solid line is the total.}
\end{figure} 
\begin{table*}
\centering
\begin{minipage}{180mm}
\caption{\label{props} Derived properties for our sample. Errors are marginalized 68\% confidence limits.}
\scriptsize
\begin{tabular}{lcc rrr rrr rrr}
Source & $z^a$ & $R_{\rm{eff}}^b$ & $T_{\rm{d}}$ & $\beta$ & $\log M_{\rm{d}}$ & $\tau_{V}$ & $\log M_*$ & $\log L_{3-1000}$ & $\log L_{\rm{PAH}}^c$ & SFR$^d$ & Quality$^e$ \\
 & & kpc &  K & & \msun\ & & \msun\ & \lsun\ & \lsun\ &  \msun/yr & \\
 & & & & & & &  & & & & \\
N1-001 & 0.030 & 4.41 & 25.2\,$\pm$\,4.7 & 1.9\,$\pm$\,0.5 & 7.10\,$\pm$\,0.22 & 0.95\,$\pm$\,10.15 & 9.44\,$\pm$\,0.09 & 10.06\,$\pm$\,0.06 & 9.07\,$\pm$\,0.21 & 1.72\,$\pm$\,0.29 & 2 \\ 
N1-002 & 0.064 & 9.55 & 21.9\,$\pm$\,6.1 & 2.1\,$\pm$\,0.6 & 7.62\,$\pm$\,0.24 & 5.41\,$\pm$\,5.60 & 10.47\,$\pm$\,0.16 & 10.62\,$\pm$\,0.06 & 10.15\,$\pm$\,0.20 & 5.3\,$\pm$\,1.1 & 2 \\ 
N1-004 & 0.064 & 14.29 & 23.0\,$\pm$\,5.6 & 2.1\,$\pm$\,0.5 & 7.56\,$\pm$\,0.19 & 3.31\,$\pm$\,3.50 & 10.43\,$\pm$\,0.13 & 10.60\,$\pm$\,0.08 & 9.71\,$\pm$\,0.24 & 5.7\,$\pm$\,1.2 & 1 \\ 
N1-007 & 0.061 & 6.10 & 24.1\,$\pm$\,7.1 & 1.4\,$\pm$\,0.7 & 7.50\,$\pm$\,0.23 & 5.40\,$\pm$\,5.03 & 10.25\,$\pm$\,0.18 & 10.42\,$\pm$\,0.11 & 9.71\,$\pm$\,0.21 & 3.6\,$\pm$\,1.0 & 2 \\ 
N1-009 & 0.053 & 5.56 & 23.3\,$\pm$\,6.6 & 2.1\,$\pm$\,0.6 & 7.18\,$\pm$\,0.32 & 2.67\,$\pm$\,2.37 & 10.32\,$\pm$\,0.13 & 10.22\,$\pm$\,0.08 & 9.61\,$\pm$\,0.18 & 2.02\,$\pm$\,0.47 & 1 \\ 
N1-012 & 0.066 & 6.88 & 27.4\,$\pm$\,8.3 & 1.7\,$\pm$\,0.6 & 7.35\,$\pm$\,0.37 & 2.27\,$\pm$\,1.68 & 9.65\,$\pm$\,0.11 & 10.44\,$\pm$\,0.13 & 9.26\,$\pm$\,0.25 & 4.2\,$\pm$\,1.4 & 1 \\ 
N1-013 & (0.46) & 18.63 & 32.5\,$\pm$\,11.6 & 2.0\,$\pm$\,0.8 & 8.73\,$\pm$\,0.34 & 7.81\,$\pm$\,4.12 & 10.79\,$\pm$\,0.16 & 12.11\,$\pm$\,0.11 & 10.64\,$\pm$\,0.23 & 210\,$\pm$\,55 & 1 \\ 
N1-015 & 0.234 & 12.37 & 23.7\,$\pm$\,5.1 & 2.1\,$\pm$\,0.5 & 8.50\,$\pm$\,0.37 & 2.85\,$\pm$\,1.63 & 10.56\,$\pm$\,0.16 & 11.39\,$\pm$\,0.12 & 10.40\,$\pm$\,0.25 & 36\,$\pm$\,13 & 1 \\ 
N1-016 & 0.092 & 6.10 & 31.8\,$\pm$\,8.0 & 1.6\,$\pm$\,0.6 & 7.44\,$\pm$\,0.35 & 1.73\,$\pm$\,8.02 & 10.25\,$\pm$\,0.15 & 10.86\,$\pm$\,0.08 & 10.29\,$\pm$\,0.26 & 11.0\,$\pm$\,1.9 & 1 \\ 
N1-024 & 0.086 & 5.44 & 24.2\,$\pm$\,7.9 & 1.8\,$\pm$\,0.7 & 7.73\,$\pm$\,0.23 & 3.15\,$\pm$\,2.71 & 9.95\,$\pm$\,0.16 & 10.60\,$\pm$\,0.09 & 9.76\,$\pm$\,0.20 & 6.1\,$\pm$\,1.6 & 1 \\ 
N1-029 & 0.144 & 9.75 & 26.6\,$\pm$\,16.9 & 2.0\,$\pm$\,0.8 & 7.88\,$\pm$\,0.59 & 3.50\,$\pm$\,2.74 & 10.40\,$\pm$\,0.14 & 10.93\,$\pm$\,0.12 & 10.09\,$\pm$\,0.23 & 12.2\,$\pm$\,4.5 & 1 \\ 
N1-031 & 0.063 & 4.29 & 21.9\,$\pm$\,15.5 & 2.4\,$\pm$\,0.8 & 7.22\,$\pm$\,0.52 & 2.08\,$\pm$\,2.03 & 9.81\,$\pm$\,0.12 & 10.16\,$\pm$\,0.13 & 9.20\,$\pm$\,0.25 & 2.07\,$\pm$\,0.69 & 1 \\ 
N1-039 & 0.269 & 14.21 & 24.5\,$\pm$\,9.4 & 2.5\,$\pm$\,0.6 & 8.44\,$\pm$\,0.70 & 7.21\,$\pm$\,5.34 & 10.53\,$\pm$\,0.22 & 11.37\,$\pm$\,0.16 & 10.21\,$\pm$\,0.24 & 35\,$\pm$\,13 & 1 \\ 
N1-040 & 0.450 & 26.34 & 26.5\,$\pm$\,7.9 & 1.8\,$\pm$\,0.7 & 9.10\,$\pm$\,0.21 & 15.61\,$\pm$\,7.36 & 10.42\,$\pm$\,0.34 & 11.83\,$\pm$\,0.13 & 9.85\,$\pm$\,0.41 & 104\,$\pm$\,33 & 1 \\ 
N1-041 & 0.120 & 7.85 & 24.7\,$\pm$\,12.2 & 2.3\,$\pm$\,0.7 & 7.82\,$\pm$\,0.77 & 4.71\,$\pm$\,2.51 & 10.01\,$\pm$\,0.14 & 10.77\,$\pm$\,0.14 & 9.93\,$\pm$\,0.23 & 8.2\,$\pm$\,3.0 & 1 \\ 
N1-045 & (0.18) & 15.74 & 28.1\,$\pm$\,14.6 & 1.1\,$\pm$\,0.8 & 8.08\,$\pm$\,0.57 & 2.55\,$\pm$\,0.00 & 10.53\,$\pm$\,0.16 & 11.08\,$\pm$\,0.13 & 10.23\,$\pm$\,0.16 & 17.4\,$\pm$\,6.0 & 1 \\ 
N1-064 & 0.910 & 17.54 & 31.1\,$\pm$\,6.3 & 1.8\,$\pm$\,0.7 & 9.32\,$\pm$\,0.21 & 29\,$\pm$\,30 & 11.47\,$\pm$\,0.25 & 12.45\,$\pm$\,0.13 & 11.73\,$\pm$\,0.39 & 410\,$\pm$\,130 & 3 \\ 
N1-068 & (0.16) & 9.69 & 26.7\,$\pm$\,20.3 & 2.1\,$\pm$\,0.8 & 7.83\,$\pm$\,0.75 & 3.41\,$\pm$\,1.99 & 9.99\,$\pm$\,0.15 & 10.93\,$\pm$\,0.14 & 9.88\,$\pm$\,0.20 & 13.0\,$\pm$\,4.4 & 1 \\ 
N1-077 & (0.20) & 11.60 & 27.1\,$\pm$\,14.5 & 2.0\,$\pm$\,0.7 & 7.91\,$\pm$\,0.65 & 3.25\,$\pm$\,2.27 & 10.17\,$\pm$\,0.15 & 10.92\,$\pm$\,0.15 & 10.06\,$\pm$\,0.21 & 12.0\,$\pm$\,4.6 & 1 \\ 
N1-078 & (0.91) & 16.33 & 69.8\,$\pm$\,15.0 & 1.1\,$\pm$\,0.5 & 9.16\,$\pm$\,0.25 & 27\,$\pm$\,30 & 11.49\,$\pm$\,0.26 & 12.90\,$\pm$\,0.22 & 11.59\,$\pm$\,0.38 & 1360\,$\pm$\,570 & 3 \\ 
N1-083 & (0.31) & 14.17 & 30.8\,$\pm$\,18.3 & 2.0\,$\pm$\,0.7 & 8.21\,$\pm$\,0.58 & 4.81\,$\pm$\,4.19 & 10.65\,$\pm$\,0.23 & 11.40\,$\pm$\,0.16 & 10.34\,$\pm$\,0.19 & 35\,$\pm$\,15 & 1 \\ 
N1-101 & 0.060 & 6.16 & 24.6\,$\pm$\,17.8 & 2.1\,$\pm$\,0.8 & 7.07\,$\pm$\,0.75 & 4.94\,$\pm$\,3.33 & 9.36\,$\pm$\,0.18 & 9.89\,$\pm$\,0.14 & 9.01\,$\pm$\,0.19 & 1.07\,$\pm$\,0.44 & 1 \\ 

\end{tabular}
\medskip

$^a$ Spectroscopic redshifts from D05 and Chapman et al.~(2002) (N1-040 and N1-064); photometric redshifts given in brackets (see Appendix~B). \\
$^b$ The observed half-light radius at 8\um\ converted to physical distance. \\
$^c$ The integral under our full PAH template.\\
$^d$ As per Kennicutt~(1998). For consistency $L_{8-1000}$ was used here.\\
$^e$ Quality flag: 1=good; 2=$\chi_{\rm{red}}^2>2$; 3=$\tau_{V}$ poorly fit (see Section A.1). \\ 
\end{minipage}
\end{table*}
\section{SED model}
\subsection{Interpreting the observed SEDs}
The observed SED of a galaxy depends on the properties of the stellar populations (e.g. ages and metallicities), the dust model (e.g. composition, size distributions), and the relative geometry of the two.  The full range of possible combinations of all these is such that, fully accounting for all possible processes is todate not possible.  Radiative transfer models where stars form inside giant molecular clouds and gradually disperse into the diffuse medium exist \citep{silva98,cirrus_mrr,popescu00}, but tend either to have too many free parameters, or not cover the full range in SEDs observed. Self-consistent models of the stars and dust (i.e. energy absorbed equals energy emitted) have been done on local galaxies \citep{galliano03}; however, require exceptionally well sampled SEDs from the UV to the sub-millimeter.  We choose a modelling approach (see Section~3.2) that is data-driven: i.e. our model is no more complicated than our data allow us to fit, yet is flexible enough to describe the full range of SEDs observed in the sample.  Such a simple phenomenological model has the effect of smoothing over the underlying messiness of galaxy SEDs, while providing us with best-fit parameters (such as opacity and dust temperature) those values are some 'characteristic' or 'average' values of an underlying distribution.  The approach is therefore well suited to broadband data, such as presented for our sample in the previous section, which is, in essence, spatially integrated over the entire galaxy, including its quiescent (i.e. cirrus) and potentially active (i.e. starburst) environments.  \\
The main limitation in interpreting our results in terms of the physical nature of our galaxies is due to the fact that, as discussed in the previous section, the available optical/UV data ({\em U}, {\em G}, and {\em R}) for our sample are quite scarce and thus here we only discuss the spectral properties of our galaxies longward of and including the near-IR ($\sim$\,2\um), which automatically precludes us from {\em directly} discussing young stars. Thus our restriction of fitting only the $\gs$\,2\um\ spectra means that: 1) we cannot balance the energy absorbed in the optical/UV with the energy re-emitted in the infrared, 2) we cannot explicitly discuss either the star-formation history or age distribution of the stellar populations of our galaxies, and 3) we cannot constrain multiple optical depths (e.g. molecular cloud and cirrus).  These limitations are largely addressed post-fitting, i.e. in the interpretation of the best-fit model parameters and how they are used to infer underlying physical characteristics. For example, we confine our stellar model to an old stellar population template (see Section~3.2), but when deriving stellar mass, we use mass-to-light ratios which include the effect of young stars on the {\em K}-band (Section~4.3).  We know that young stars form inside dense molecular clouds and later disperse into the diffuse 'disk' environment. Thus the characteristic optical depth absorbing the young stars' power is likely not the optical depth we derive in the near-IR which is dominated by the older stellar populations.  We address this in Section~5.2, where the dust masses are derived both from emission (predominantly young stars) and absorption (predominantly old stars in our case).  \\
 With the above philosophy in mind, we proceed with the model procedure outlined below. 
\subsection{Details of Model Procedure}
\label{secmodel}
As discussed above, our modelling approach is driven by the desire to have a phenomenological model which fits known nearby galaxy SEDs robustly and yet has the flexibility (i.e. number of phenomenological parameters only slightly less than the number of data points we are trying to fit) which allows us to fit a wide range of SED types, and in addition we want to be able to extract physical parameters with meaningful uncertainties (in particular such that degeneracies between parameters can be highlighted). With these goals clearly in mind, our choice of detailed SED model is dependent on the quality and wavelength coverage of the data, while the statistical approach we use is the Markov Chain Monte Carlo method.\\
We model the SED as a sum of stellar emission, PAH emission, power-law emission, and thermal grey-body emission. This is based on the mid-IR model used in Sajina, Lacy, and Scott~(2005). \\
The stellar emission is accounted for by a 10\,Gyr-old, solar metallicity, Salpeter IMF, single stellar population (SSP) spectral template generated with the {\sc PEGASE2.0} spectral synthesis code (Fioc et al.~1997).  The specific stellar model used here is not important, being merely a stand-in for 'old stars'. For any population older than about 1Gyr,  the near-IR spectrum looks essentially the same (through a couple of broadband filters), the differences being in the shorter wavelengths. Since our sources are largely $z$\,$\sim$\,0, with a tail extending up to $z$\,$\sim$\,1, this can be thought of as: all of our sources include {\em some} stars that are as old as the Universe at the observed epoch. \\
We use the neutral PAH emission template of Lee \& Draine (2003). The only modification we make is the addition of the newly discovered 17\um\ feature (Smith et al.~2004). Our broadband data do not allow us to investigate the probable variations in relative PAH feature strength, and therefore any reasonable PAH template can be used here, with the understanding that it is only an approximiation for any given galaxy. A power-law, $f_{\nu}$\,$\propto$\,$\nu^{-\alpha}$, is a proxy for the warm, small grain emission (roughly $<$\,60\um). This is cut-off as $\exp(-0.17$\,$\times$\,$10^{14}\,\rm{Hz}/\nu)$ in order not to interfere with the far-IR/sub-mm wavelength emission, which is described by a thermal, $f_{\nu}$\,$\propto$\,$\nu^{3+\beta}[\exp(h\nu/kT)-1]^{-1}$, component. Since we typically only have the 24\um\ point to constrain this component, we fix $\alpha$ at 3.0, which smoothly connects this to the greybody component.  Extinction, parameterized by $\tau_{\rm{V}}$, is applied using the $R_{\rm{V}}$\,=\,3.1 Milky Way-type extinction curve of Draine (2003). This includes the 9.7\um\ Si absorption feature, which thus becomes noticeable in this model at high opacities. We consider only a screen geometry, i.e. $I_{\nu}$\,=\,$I_{0}\exp(-\tau_{\nu})$. Fig.~\ref{sed_demo} shows the above phenomenological break-up of the SED. We solve for the best-fit model, and the associated uncertainties in the parameters via Markov Chain Monte Carlo (MCMC). Details of the fitting procedure and error analysis are given in Appendix~A. \\
This model is complicated by the lack of spectroscopic redshifts for about 1/3 of the sample. We discuss our approach to determining photometric redshifts for these sources in Appendix~B. 
\begin{figure*}
\centering
\vspace*{14cm}
\leavevmode
\includegraphics{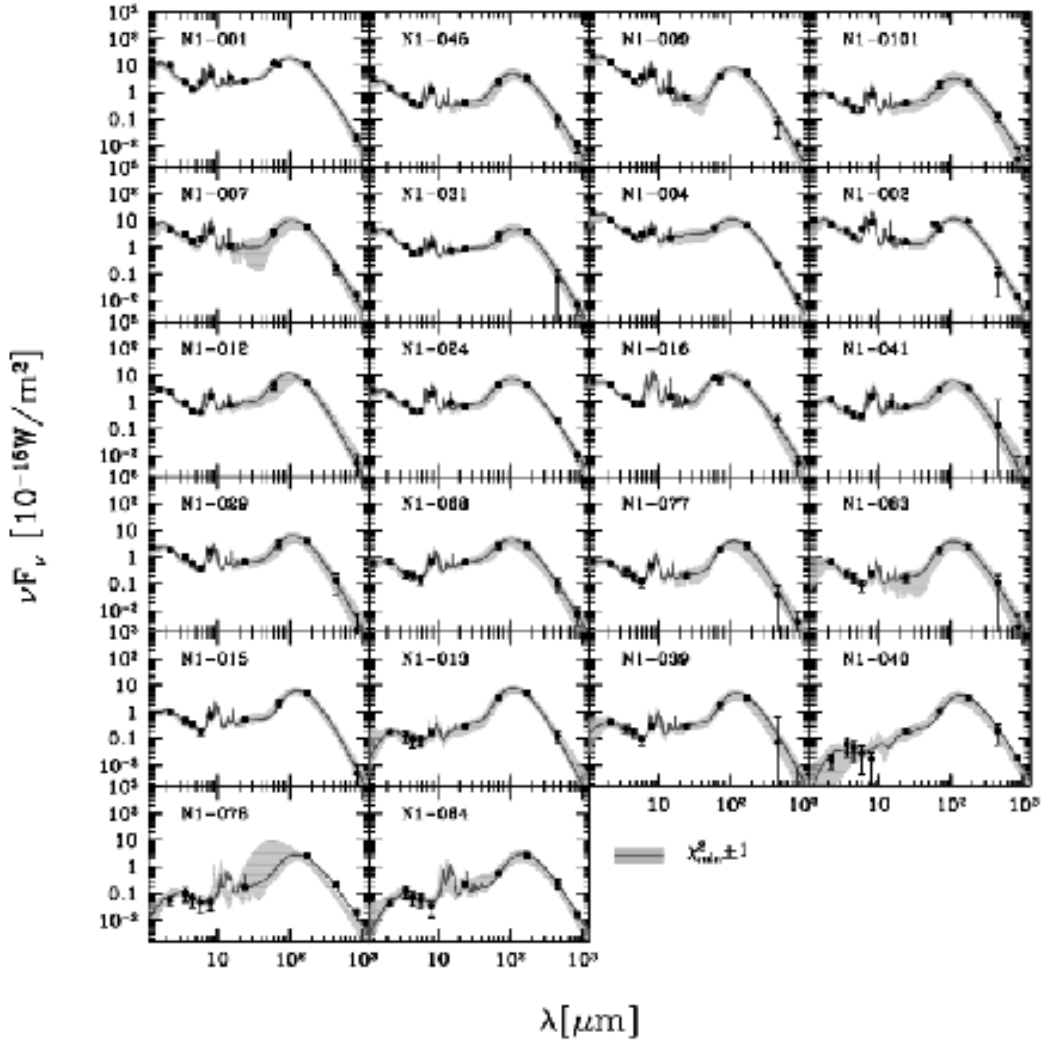}
\caption{\label{seds} The SED fits for our sample, where we show both the data and best-fit model together with an uncertainty band. Note that the least well-defined SED is that of N1-078, which is the only source without data in the range 24\um\,--\,170\um.  The sources are ordered so that redshift increases from top to bottom. Note the characteristic increase in the dust-to-stars emission ratio as one goes to higher redshift (and hence more luminous) sources.}
\end{figure*}    
\section{Fit results}
\label{secresults}
The quantitative conclusions of our model-fitting are presented in Table~\ref{props}, where the errors represent marginalized 68\% confidence limits. A quality flag is also given to indicate whether or not problems exist with the fit. Below we discuss both the general aspects of the fitted SEDs, and details of the parameters presented in Table~\ref{props}.
\subsection{General trends in the SEDs}
\label{secseds}
Fig.~\ref{seds} shows the data overlaid with the best-fit SED model for sources ranging between $z$\,$\sim$\,0 (top row) and $z$\,$\sim$\,1 (bottom row).  
As expected, there is a greater range in SED shapes than accounted for by the data uncertainties. However, there are natural groupings, for example one can easily distinguish ULIGs (e.g.~N1-064, N1-078) from LIGs (N1-015, N1-039) and from normal galaxies (e.g. N1-002, N1-009). The grey shading in Fig.~\ref{seds} gives a sense of the uncertainty of the SED fits. Without prior assumptions on the shape of the SED (s.a. by using templates) as in this case, if we do not have constraints on both sides of the thermal peak (e.g. N1-078), a wide range of SED shapes and consequently temperatures, luminosities etc. are acceptable.   

\begin{figure}
\centering
\vspace*{8cm}
\leavevmode
\includegraphics{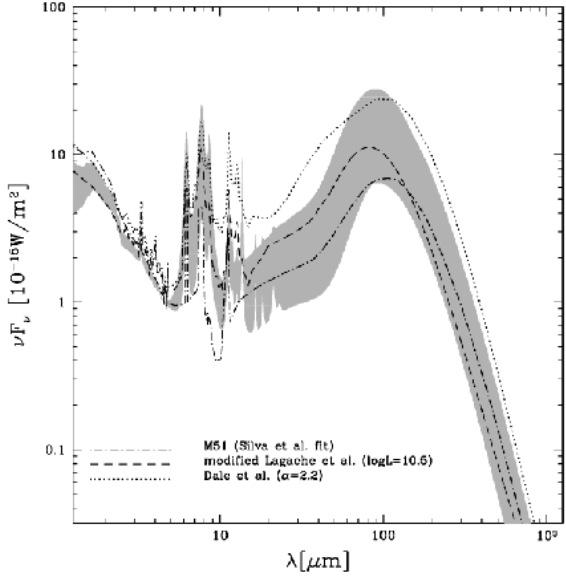}
\caption{\label{coldspec} The shaded region represents the range of spectra, where we have included all sources with $z$\,$<$0.1 and which have 24\um\ data (except N1-009, which has an unusually strong stellar component, for our sample). All SEDs are normalized at 4.5\um, which is a neutral point between pure stellar and PAH emission. The shaded region should be regarded as a composite spectrum, representative of the `cold' FIRBACK sources. For comparison we overlay a number of templates (see legend). }
\end{figure}

In Fig.~\ref{coldspec}, we address the question of what is the `typical' normal/cold galaxy spectrum based on our model fits. In order to minimize redshift bias, we construct a composite spectrum from all sources with redshifts below 0.1, and for which 24\um\ data are available. To minimize luminosity effects, we also normalize the spectra at 4.5\um\ (which point resulted in the least amount of scatter across the SED) and effectively is a normalization in stellar mass. This procedure is consistent with results for {\sl ISO} Key Project normal galaxies, where the stellar+PAH SEDs were found to be fairly constant \citep{lu03}. Comparing the results with a number of common SED models, we find that our sample is best described by fairly cold spectra with low mid-IR/far-IR ratios. 
In Fig.~\ref{seds}, we see that the LIGs (i.e.~presumably starbursts) also have fairly cold spectra, unlike that of the `prototypical' starburst M82 (this was already noted in D05).              

\begin{figure}
\centering
\vspace*{8cm}
\leavevmode
\includegraphics{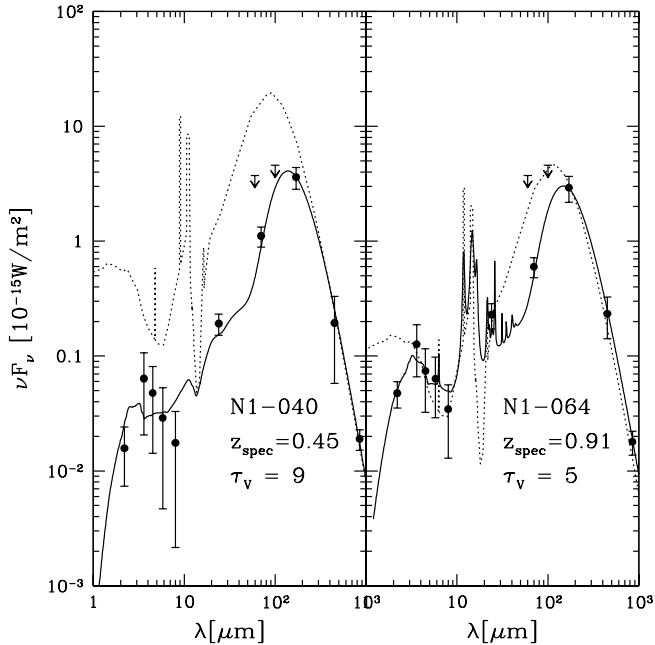}
\caption{\label{uligs} The best-fit SEDs for the two spectroscopically-confirmed higher-$z$ sources. The {\sl IRAS} upper limits are also shown, although they are not included in the fit. The dotted line shows the appropriately redshifted Arp220 template \citep{silva98}. The {\sl Spitzer} fluxes suggest cooler and less luminous sources than previously assumed. 
}
\end{figure}

Even more surprisingly, this cold trend does not appear to reverse for the highest luminosity sources including ULIGs. It must be noted however that it is not clear how common are such SEDs for ULIGs in general as we only have two spectroscopic ULIGs in our sample and our far-IR selection likely biases us toward colder sources. In Fig.~\ref{uligs} we show the best-fit SEDs for the two spectroscopically confirmed $z$\,$\sim$\,0.5\,--\,1.0 sources \citep{c02}. The dotted line is an Arp220 template \citep{silva98} at the appropriate redshift. In both cases the sub-mm data are well fit by the Arp220 template, but the mid-IR data differ by an order of magnitude. Our photometric highest-$z$ source (N1-78) shows similar trends. Note that the PAH features in the best-fit N1-064 spectrum are not constrained by any data point and therefore a model with no PAH emission is quite acceptable as well (see the spread in Fig.~\ref{seds}). Despite it being poorly constrained for the few highest-$z$ sources, we find clear evidence for prominent PAH emission for nearly all sources in our sample, regardless of redshift (and hence luminosity).
  
The weak mid-IR continua, and strong PAH emission of our sources suggest that they are star-formation rather than AGN dominated (see e.g.~Sajina, Lacy, Scott 2005).
\subsection{Luminosity and SFR}
\label{seclum}
In Table~\ref{props}, we present both the total infrared power output of our sources ($L_{3-1000}$), as well as the luminosity due to PAH emission alone. 

The overall luminosities we derive are typically a few\,$\times$\,$10^{10}$\lsun\ for our lower-$z$ targets, consistent with roughly $L^*$ galaxies. $L_{\rm{PAH}}$ is obtained by integrating under the PAH component alone. We find the $L_{\rm{PAH}}/L_{\rm{IR}}$ fraction to be typically 5\,--\,15\%. This is consistent with the results for the {\sl ISO} Key Project galaxies \citep{dale01}. 

\begin{figure}
\centering
\vspace*{8cm}
\leavevmode
\includegraphics{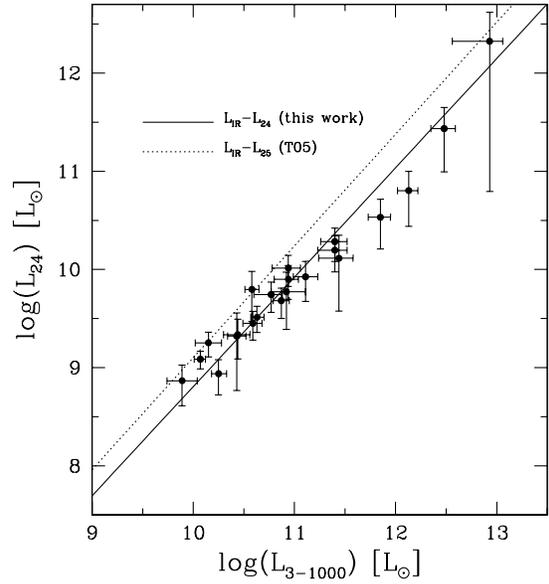}
\caption{\label{lums}  The 24\um\ luminosity vs. total infrared luminosity. Note that within the uncertainties we find no difference between the {\sl IRAS} 25\um\ luminosity and the MIPS 24\um\ luminosity. See Section~\ref{seclum} for details.}
\end{figure}

Since such well-sampled SEDs are rare, to obtain the overall infrared luminosity, extrapolations from the mid-IR are common. However, the coldness of our sources means that, if such sources are the norm, prior relations might not be applicable. In Fig.~\ref{lums} we compare our results on the mid-IR/total-IR relation with previous, {\sl IRAS}-based, results (e.g.~Takeuchi et al.~2005). The solid line is the best-fit for our sample, which is:
\begin{eqnarray}
\log L_{24}=(1.13\pm0.05)\times \log L_{\rm{IR}}-(2.5\pm0.5).
\end{eqnarray} 
Our rest-frame 24\um\ fluxes are below those expected from the {\sl IRAS} relation by $\sim$\,0.4\,dex. Although we have too few sources at the high-$L$ end to claim this conclusively, it appears that the relation drops even further for these sources. More likely, however, this is just an indication of the underlying scatter in the relationship due to variations in the SED shape.    

This discrepancy with the earlier relation is most likely a selection bias. The sample used for the {\sl IRAS} relation is flux-limited to all four {\sl IRAS} bands, leading to a bias toward sources with stronger warm continuum, unlike our 170\um\ selection where the bias is toward the presence of cold dust instead. We return to this point in Section~5.1.

About 70\% of the sample have $\log(L/L_{\odot})$\,$<$\,11, while the remaining 30\% have $\log(L/L_{\odot})$\,$>$\,11. This can be thought of as the break-up between fairly quiescent and more actively starforming galaxies (we address the mode of star-formation of our galaxies in detail in Section~5.3).
The total infrared luminosity is well known to trace the current SFR of a galaxy \citep[see e.g.][]{k98,kewley02}. We use the Kennicutt relation here which is:
\begin{equation}
\frac{\rm{SFR}}{\rm{M}_{\odot}\rm{yr}^{-1}} = (1.8\times10^{-10})\frac{L_{8-1000}}{\rm{L}_{\odot}},
\end{equation}
Note the slightly different definition of $L_{\rm{IR}}$, which is accounted for here (i.e.~8\,--\,1000\um\ rather than 3\,--\,1000\um). Typically our galaxies have SFR of $\sim$\,5\,$M_{\odot}$/yr, which is slightly enhanced with respect to local quiescent spirals. As expected, the SFRs rise to a few hundred $M_{\odot}$/yr for our few ULIGs. 
  
\subsection{Stellar mass and dust obscuration}
\label{secstars}
As part of our SED model fitting we estimate the $k$-corrected, and dust corrected near-IR stellar emission. The near-IR is preferred for deriving stellar mass as it is fairly robust against both uncertainties in the dust obscuration level (a factor of $\sim$\,10 less so than in the optical), and to the details of the stellar population and SFH (being largely sensitive to the red giant population only). Without sufficient optical coverage, we cannot properly account for the possible contribution of young stars. We merely calculate the un-extincted, restframe $L_{\rm{K}}$, and convert this to stellar mass by assuming an appropriate {\em K}-band mass-to-light ratio, $\gamma_{K}$. Despite its relative insensitivity to young stars, the near-IR mass-to-light ratio does increase as the fraction of young/massive stars in the galaxy increases. In addition, a degeneracy exists in that a more massive and dustier galaxy can appear similar to a less massive and less obscured one. The significance of this effect grows for sources without {\em J}-band data or with poor near-IR SNR, both of which are more common for the higher-$z$ sources. However, since the optical depth is a free parameter in our model, it is already accounted for in the resulting Markov chains. The uncertainty in mass-to-light ratio however is not included. The additional uncertainty on the stellar mass is: $d\log M_{*}$\,=\,$d\log\gamma$. A mean value of $\gamma_{K}$\,=\,0.6 was found by \citet{bdj01} for their sample of spiral galaxies, while \citet{gdp00} find $\gamma_{K}$ to be $\sim$\,0.92 for their sample of starburst and H{\sc ii} galaxies. \citet{bdj01} consider a wide range of reasonable population models, finding $d\log\gamma_{K}$\,$\sim$\,0.2\,dex (although this is smaller if the {\em B}\,--\,{\em K} colour is known, due to the effect of the young stars). The SFRs derived for our sources in the previous section suggest enhanced star-formation activity compared with normal spirals. Therefore, the Gil de Paz et al value of $\gamma_{K}$\,=\,0.92 is likely more appropriate for our sample, while we assume the Bell\,\&\,de Jong uncertainty (from the unknown SFHs) of 0.2. Adding this in quadrature to the MCMC derived errors suggests that more realistic uncertainties here are about twice the quoted ones. 

With the above set of assumptions, we find $\langle M_{\rm{star}} \rangle $\,$\sim$\,2\,$\times$\,10$^{10}$\msun,\footnote{For comparison, typical $M^*$ values are $\sim$\,3\,$\times$\,10$^{10}$\msun.} with the $\log(L/L_{\odot})$\,$<$\,11 sources and those with $\log(L/L_{\odot})$\,$>$\,11 differing by $\sim$\,0.6\,dex. Note, however, that from the above discussion we expect that somewhat higher values for $\gamma$ are appropriate for the more luminous (higher SFR) galaxies, and vice versa for the lower luminosity ones. Taking this into account, the observed difference in mass may be even largely in reality. 

For most sources, we find modest levels of dust extinction (0\,$<$\,$\tau_{V}$\,$<$\,5, peaking at $\sim$\,3) for most sources. This assumes a screen geometry, and would increase in a uniform mixture of dust and stars. The amount of extinction generally increases for $z$\,$\gs$\,0.3 sources, with N1-040 requiring the greatest optical depth, consistent with its near-IR faintness (note however that the mean-likelihood and MCMC approaches disagree on the optical depths of the high-$z$ sources -- see Appendix~A).  

The values of $\tau_{V}$ we find  for the bulk of the FIRBACK N1 galaxies, is consistent with the average $\tau_{V}$\,$\sim$\,3 found from the $H_{\alpha}/H_{\beta}$  ratio of the FIRBACK-South galaxies \citep{patris}. This means that, typically, the optical depths to which the light of old and young stars is subjected do not differ dramatically when integrated across the galaxy (see also Section~5.2).

\subsection{Dust properties: $T$\,--\,$\beta$ relation and dust mass}
\label{secdust}

The single greybody approach we take (see Section~\ref{secmodel}) in modeling the dust emission of these galaxies, although in common use and the only one possible when just a few data points are available, is clearly an approximation \citep[see also][]{de01,blain_seds}. More realistically, a distribution of dust grain characteristics, such as optical properties and geometry, results in a distribution of effective temperatures and $\beta$'s.  In addition, degeneracies arise in fitting this model due to the functional form and spectral sampling (see Appendix~\ref{secgrey} for further discussion).
\begin{figure}
\label{betat}
\centering
\vspace*{4.2cm}
\leavevmode
\includegraphics{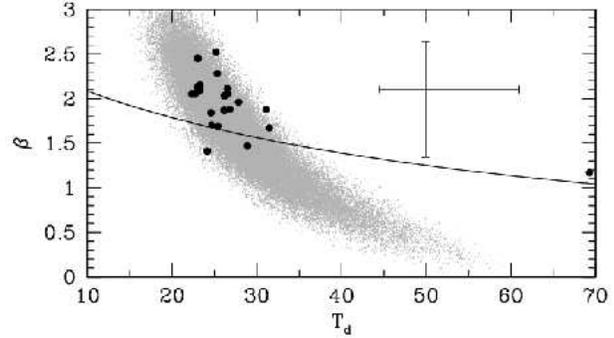}
\caption{The best fit values (black dots) for \td\ and $\beta$. The solid line is the \citet{dupac03} relation (see Appendix~C). For comparison the greyscale points show the $\chi^2_{\rm{min}}$\,+\,3 region for N1-024. The error bars represent the average 68\% uncertainties on individual sources (note that it is driven by the least constrained sources, while the grey points give a better sense of the scatter in well constrained sources). The outlier near 70\,K is N1-078, for which no data exist between 24\um\ and 170\um.
}
\end{figure}

For our sample, values of $T_{\rm{d}}$\,$\simeq$\,20\,--\,30\,K and high values of $\beta$ ($\simeq$\,2) are most representative. These are in good agreement with the typical far-IR temperatures ($\sim$\,22\,K) of local spiral and irregular galaxies \citep{contursi01,serendipity}, as well as theoretical expectations which place the big grain emission (dominating the thermal peak of normal galaxies) of a standard ISM at \td\,$\sim$\,15\,--\,30\,K with $\beta$\,$\sim$\,2 \citep{dl84}. Our estimates are also consistent with, but somewhat cooler than, the values found by \citet{taylor05} for the FIRBACK galaxies in ELAIS-N2.   

The cold dust mass is estimated in the usual way: $M_{\rm{d}}$\,=\,$S_{850}*D_{L}^2[(1+z)\kappa B(T,\nu_{e})]^{-1}$ \citep[e.g.][]{duncan}, where $B(T,\nu_{e})$ is the Planck function for the given temperature, at the emission frequency, and the dust absorption coefficient $\kappa$ parameterizes the unknown grain properties. We assume the value of $\kappa$=0.077\,$\pm$\,0.02\,$\rm{m}^2\rm{kg}^{-1}$, as advocated by \citet{james03}, although there are arguments for a value up to 3 times higher \citep{dasyra}. We return to the effect of increasing $\kappa$ in Section~5.2. We find $M_{\rm{d}}$\,$\sim$\,$10^{7}$--$10^{8}$\,$M_{\odot}$, which is comparable with the masses found for the SCUBA Local Universe Galaxies Survey (SLUGS) galaxies \citep{de01}. As expected, the higher $z$, more luminous galaxies have higher dust masses, typically $\sim$\,10$^8$\,--\,10$^9$\msun. 

\subsection{Size and morphology}
\label{secsize}

The closest galaxies in our sample appear disk-like by visual inspection of the IRAC images (see Fig.~\ref{mosaic}). In addition, we would like to determine the typical sizes of our sample, and whether there is a noticeable morphological difference as a function of infrared activity (i.e.~$L_{\rm{IR}}$). Fortunately, all 22 sources considered here are at least mildly resolved by IRAC (diffraction limit $\sim$\,2\arcsec\ at 8\um). The effective radii given in Table~\ref{props} are the 8\um\ half-light radii converted to physical distance (note there is no noticeable difference here if we used any other IRAC band). We find that, typically, $R_{\rm{eff}}$\,$\sim$\,5\,--\,10\,kpc, although a tail of the sample extends to larger radii. These sizes are consistent with those of large spirals.

\section{Analysis}
In this section, we use the results of the previous section in attempting to arrive at a consistent picture of the physical properties of our sources. We particularly focus on three related  questions. Why are all our galaxies, including the ULIGs, colder than expected? Can we say something about the spatial distribution of the star-formation activity -- i.e. is there indication of centrally concentrated starbursts surrounded by older stellar haloes, or is the star-formation by contrast distributed throughout the galaxy? And finally, is the typical FIRBACK galaxy actually starbursting?. 
\subsection{Why cold LIGs and ULIGs?}
\begin{figure}
\centering
\vspace*{8cm}
\leavevmode
\includegraphics{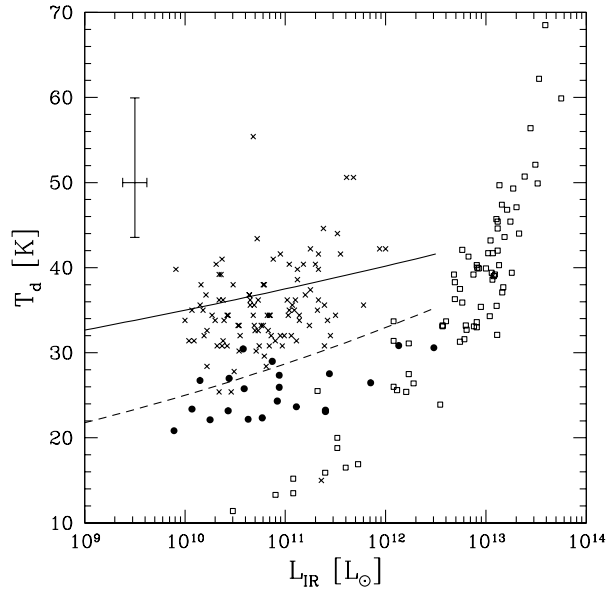}
\caption{\label{lum_td} A comparison of the luminosity-temperature relation for our sample ({\it solid circles}) with other infrared-selected galaxies including {\sl IRAS}-bright galaxies \citep{de01} ({\it crosses}), and SCUBA-selected galaxies \citep{c05} ({\it open squares}). The solid and dashed lines represent the loci for merging and quiescent galaxies respectively \citep{barnard}. Note that our sample is consistent with the expectation for quiescent starformers. }
\end{figure}

Both our general inspection of the SEDs and the derived dust temperatures for our sample suggest that fairly cold spectra  with weak (compared with {\sl IRAS} galaxies) mid-IR emission describe our entire sample, despite three orders of magnitude variation in bolometric luminosity. This can be seen directly by examining the $L$\,--\,$T_{\rm{d}}$ relation \citep[see][]{blain_seds}. This relation is of particular significance to far-IR/sub-mm selected samples, as in principle it can reveal something of the nature of the sources based on only a few spectral points (enough, for example, for a single greybody fit), because it is simply a relation between the location of the far-IR peak and its overall strength. In Fig.~\ref{lum_td}, we compare the location of our sample in the ($L$,~$T_{\rm{d}}$) plane with other infrared-selected samples, namely the SCUBA-selected high-$z$ sources \citep{c05} and the bright {\sl IRAS}-selected galaxies \citep{de01}. We overplot empirical relations appropriate for merging and quiescent galaxies \citep{barnard}. These can be understood intuitively by recalling that ideally, $L$\,$\propto$\,$R^2T^4$. This leads to luminosity increasing with temperature if the size is kept constant, or conversely to a more concentrated starburst being hotter than a galaxy of the same luminosity but with more spread-out star-formation (as in throughout the full disk). With a few exceptions, our sample is consistent (within the uncertainties, and given the different selection from that of {\sl IRAS} galaxies) with the relation for quiescent galaxies \citep[for further discussion of the relation see][]{scott_bivar}, and consistent with our size results in Section~4.5. 

The above discussion, taken a step further, suggests that our cold ULIGs might show extended star-formation activity, rather than the usual nuclear-concentration seen in the late stages of major mergers \citep{vk02}.  Such a scenario has in particular been proposed for some SCUBA galaxies \citep{kaviani03,cirrus_mrr}. Locally, only nuclear starbursts appear capable of reaching the requisite SFR/luminosity. However, at high-$z$, where presumably the progenitors are less evolved and more gas rich (i.e. with lower stability thresholds), it is possible that a major merger results in a disk-wide starburst (or rather many pockets throughout the disk) instead \citep{mihos99}. Alternatively, a cause of cold ULIGs might be extreme opacity around a nuclear starburst and/or AGN. 
\subsection{The spatial scales of the old and young stars}
\label{secscale}
In the previous section, we argue that the coldness of our sources, including both LIGs and ULIGs, might be due to star-formation spread over an extended cold disk, rather than a nuclear starburst surrounded by an old stellar halo. Here we apply two independent tests of this scenario. 

Our first test is based on the assumption that the IR emission is dominated by power absorbed by the dust from young stars, while the dust responsible for the near-IR opacity traces the old stellar population. The two should match in a disk-wide star-formation picture, while there is no reason for a correlation in a nuclear starburst.  In Section~4.4 we derived the dust mass based on the far-IR emission of our galaxies. In principle, the optical depth derived in Section~4.3 can also be converted to dust mass, provided the surface area of the optical region is known. We do this using the following equation, adapted from \citet{lisen_dwarfs} by using $\tau_B$\,=\,1.3\,$\tau_V$ and assuming the geometry is an inclined disk:
\begin{equation}
\label{eqtau}
M^{\rm{opt}}_{\rm{d}}=(1.1\times10^{5})\tau_{V}R^2\rm{cos} i,
\end{equation}  
where $i$ is the the inclination, such that 0$^{\circ}$ is face-on and 90$^{\circ}$ is edge-on.
 
\begin{figure}
\centering
\vspace*{8cm}
\leavevmode
\includegraphics{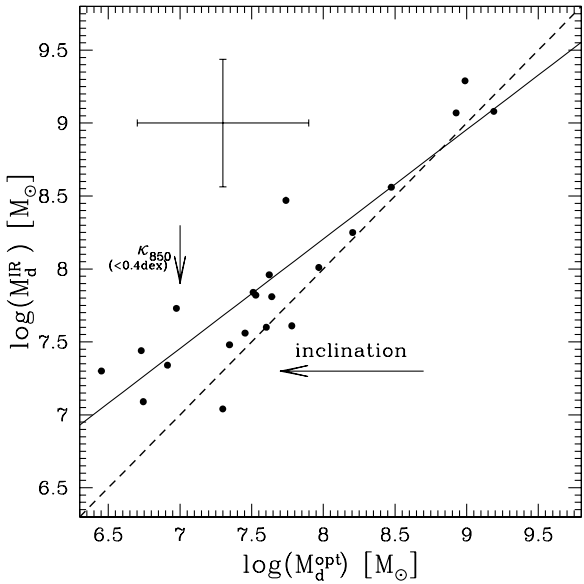}
\caption{\label{tau_md} Optically-derived dust mass (from absorption) compared with IR-derived values (from emission). The solid line is the best-fit, while the dashed line is the expected $y=x$ line (for face-on disks). The directions of increasing disk inclination, and $\kappa$ are shown. }
\end{figure}

In Section~4.5, we presented the effective radii of our sources as derived from the IRAC images. We found no difference between the 4 IRAC bands and therefore it is safe to assume that these radii represent the spatial extent of the old stellar population. In Fig.~\ref{tau_md}, we compare this optically derived dust mass and the IR-derived dust masses. The two clearly correlate, as expected. However, the best-fit slope is 0.86 unlike the expected 1.00. The arrows define the directions in which either increasing disk inclination or increasing $\kappa$ push the points. Note that increasing inclination is equivalent to decreasing the size of the region contributing the bulk of the IR emission (i.e. towards nuclear starburst). Therefore at the high-mass end, the good agreement suggests that the dust (behind the IR emission in particular) is distributed throughout the entire disk rather than being concentrated in the nuclear region (see next section). Somewhat contradictory, the high values of $\tau_{V}$ we find are also fully consistent (but the lower values implied by the likelihood analysis are not -- see Appendix~A). The two can be reconciled by a more complex geometry. 

At the low mass end we find consistently higher dust masses than expected from the above formula and the derived optical depths (hence the shallower slope). This may merely indicate that the highest-SFR regions responsible for the dust emission are behind a foreground screen of much less obscured older stars which are responsible for the near-IR emission. Another interpretation, is that our IR-based dust mass is overestimated. Adopting the extreme end of $\kappa$ values in the literature \citep{dasyra}, $\kappa$\,$\sim$\,0.2, will decrease our estimates by about 0.4\,dex, making the two estimates much more consistent. The inclination cannot reconcile them as it only affects the points to the right of the dashed line (face-on case).  This relation should probably not be over interpreted, since both expressions are approximations to much more complex systems and seeking a perfect agreement between them is therefore unrealistic. Within the uncertainties, we are gratified that our two independent estimates of the dust content agree as well as they do.  

Given the above uncertainties, and especially if $\kappa$ is underestimated, it is still possible that some of our sources (especially among the higher-$L$ ones) are primarily powered by nuclear starbursts. We can determine the actual sizes of the star-forming/bursting regions in our galaxies by using the well known relation between gas surface density and SFR surface density \citep{schmidt59}. The implied universality of the star-formation efficiency has been demonstrated observationally over many orders of magnitude in SFR density \citep{k98}. The slope of the relation SFR\,$\propto$\,$\rho^n$ is $n$\,=\,1.4 \citep[the Kennicutt relation;][]{k98}. However, the dependence of both quantities on radius is the same (since SFR and gas mass are derived from integrated properties, their respective surface densities are simply $\propto$\,$1/r^2$), meaning that a change in the effective radius takes the form of a translation along a line of slope $n$\,=\,1.0. Thus, assuming that the efficiency of star-formation is indeed universal, any departure from the Kennicutt relation could be attributed to a difference in effective radius. 
\begin{figure}
\centering
\vspace*{8cm}
\leavevmode
\includegraphics{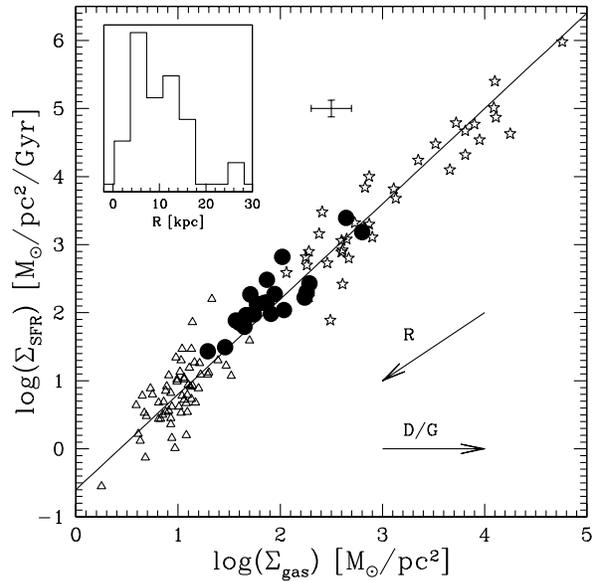}
\caption{\label{gas_sfr} The Schmidt law, $\Sigma_{\rm{SFR}}$\,=\,$\alpha$\,$\Sigma_{\rm{gas}}^n$, where observationally $n$\,=\,1.4 \citep[][solid line]{k98}. Our galaxies are plotted as the filled circles. The errorbars represent their average 1\,$\sigma$ uncertainties. For comparison we overplot the Kennicutt spiral galaxies (open triangles), and starburst galaxies (open stars) samples. The inset shows the size distribution for our sample. The slope\,=\,1 arrow shows the direction of increasing radius, while the horizontal arrow shows the direction of increasing dust-to-gas ratio. }
\end{figure} 
   
We estimate the gas mass by assuming the Milky Way gas-to-dust ratio of 100 \citep{gas-to-dust}, although the derived values, for external galaxies especially, show a large spread of about an order of magnitude \citep[see e.g.][]{serendipity}. With these estimates, in Fig.~\ref{gas_sfr}, we overplot our sources onto the original Kennicutt sample of normal spirals, and IR-bright starbursts. Most of our galaxies lie on this relation (within the errors) somewhat above the spirals, consistent with the SFRs found in Section~4.2. This suggests that the bulk of our sample is indeed forming stars throughout their extended disk, although slightly more actively (due to more gas-rich disks) than in local spirals. \\
The good agreement between our sample and the Kennicutt relation suggests that both the estimated radii and the assumed standard dust-to-gas ratio are approximately correct. The inset in Fig.~\ref{gas_sfr} gives the size distribution of the sample, which is roughly consistent with that of local spiral galaxies \citep{spiral_sizes}. An {\sl HST}-based study of the size distributions of slightly higher redshift ($z$\,$\sim$\,0.2\,--\,1) disk galaxies suggests effective radii in the same range with the peak moving from $\sim$\,6\,kpc to $\sim$\,4\,kpc as the redshift decreases \citep{goods_sizes}.     

As a final test, we double check the above conclusions by looking at the relation between $L_{\rm{IR}}$ and $T^{4+\beta}R^2$ (which should scale with luminosity, for simple geometries). As expected we find a good correlation between the two. The linear fit gives: 
\begin{equation}
\log L=(0.9\pm0.1)\times\log(T^{4+\beta}R^2)+(1.5\pm1.3),
\end{equation}
where $L$ is in $L_{\odot}$, $T$ is in K, and $R$ is in kpc. There are no obvious outliers (the rms is 0.4), again suggesting that given the cool temperatures derived, the radii inferred from the 8\um\ images are appropriate (to within a factor of 2). 

\subsection{Starbursts? It's about timescales}
In the previous section, we found that the typical FIRBACK galaxy has modest SFR, typically $\sim$\,5\msun$\rm{yr}^{-1}$, but this is somewhat enhanced activity compared with local quiescent spiral galaxies. Our investigations so far suggest disk-like star-formation, with no indication of mergers, especially for the low-$z$ sample. Now we would like to address the question of whether this enhanced star-formation could reasonably be described as a `starburst'. There is no single absolute definition of a starburst, but some commonly used indicators \citep{heckman05} include: a high intensity of star-formation; a high birthrate parameter (SFR/$\langle\rm{SFR}\rangle$); or a low ratio of the gas depletion timescale to the dynamical timescale. The question is -- when is `high' high enough, and when is `low' low enough. We now discuss each criterion in turn.

The intensity of star-formation, was indirectly addressed in Fig.~\ref{gas_sfr}. Our sources appear to fall somewhere in-between quiescent spirals and nuclear starbursts (the Kennicutt et al. starburst sample are all nuclear). Here the highest-$z$ sources (N1-064, and N1-078) are apparently the only unambiguous starbursts. The region occupied by the bulk of our {\it entire} galaxies is in fact also the locus of the {\it centres} of disk galaxies alone \citep{k98}. So, is the star-formation in our galaxies fairly smoothly distributed across the disk or in a collection of bursting knots sprinkled throughout? Are we diluting our results by assuming a smooth disk here? It appears that the intensity argument is inconclusive when applied to unresolved galaxies except in the most extreme cases.       
\begin{figure}
\centering
\vspace*{8cm}
\leavevmode
\includegraphics{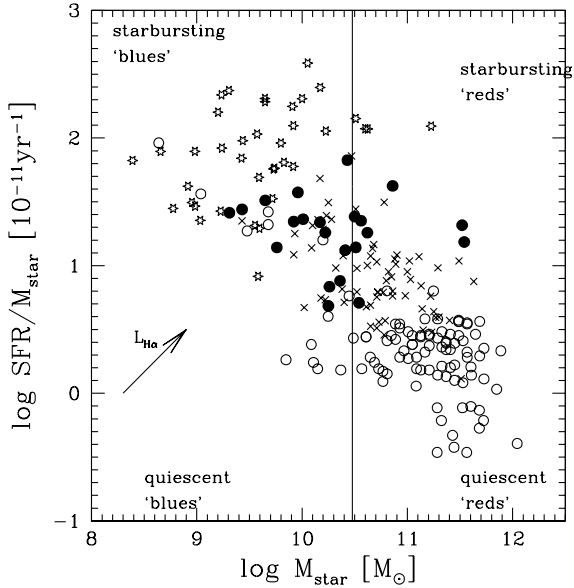}
\caption{\label{smass} Stellar mass vs.~specific SFR. This compilation of comparison galaxies is from \citet{gdp00}; however we have converted their SFRs to the standard Kennicutt et al.~relation \citep{ken_review} (a difference of --0.87\,dex). Shown are: normal spirals ({\it open circles}); local H${\alpha}$-selected star-forming galaxies ({\it crosses}); dwarf/H{\sc ii} galaxies ({\it stars}). Our FIRBACK sample is shown as the filled circles. Note that for consistency, we have converted our IR-based SFR to H${\alpha}$-based SFR using the \citet{kewley02} relation. The arrow shows the direction of increasing H${\alpha}$ luminosity from \citet{gdp00}. The mass boundary (shown by the vertical line) is the observed boundary in physical properties seen in the SDSS data \citep{kauff03a}.
}
\end{figure}

The birthrate parameter is slightly misleading, because of its dependence on the epoch of observation (although this is not too severe for our sample). A related quantity, which we examine next, is the specific SFR \citep[see e.g.][]{gdp00}, which is equivalent to the birthrate assuming a uniform epoch. The best way to interpret this is through comparison with other samples whose properties are already known, and for this purpose we use the compilation of \citet{gdp00}. However, such comparisons should be carried out with some caution, since the samples are selected in very different ways and their quoted SFRs are based on different indicators. The main distinction here is that these other samples are based on the strength of the H$\alpha$ emission line, while ours are based on the overall IR emission. The two are comparable, but not identical (largely due to the effects of dust extinction on the line strength at the high-$L$ end and possible contribution of older stars to the IR emission at low luminosities). Here we use the empirical relation of \citet{kewley02} to convert our IR-based SFRs to H$\alpha$-based ones. Lastly, we standardize the Gil de Paz et al. SFRs to the Kennicutt relation \citep{ken_review}. In Fig.~\ref{smass}, we plot stellar mass vs.~specific SFR for our sample. For comparison we overplot the data for local normal spirals, strong H$\alpha$ emitters, and H{\sc ii} dwarfs, taken from \citet{gdp00}. Fig.~\ref{smass} suggests that, overall, our sample is somewhat less massive and more active than typical local spiral galaxies. However, the bulk are larger and less star-bursting than the average local H{\sc ii} dwarf. In  fact they most strongly resemble the comparison sample of H$\alpha$-selected local star-forming galaxies \citep{gdp00}. This is to be expected, as the available spectra of our sources all show prominent emission lines (D05). The two outliers are the two ULIGs N1-064, and N1-078 which are predictably both massive and active. 
      
\begin{figure}
\centering
\vspace*{8cm}
\leavevmode
\includegraphics{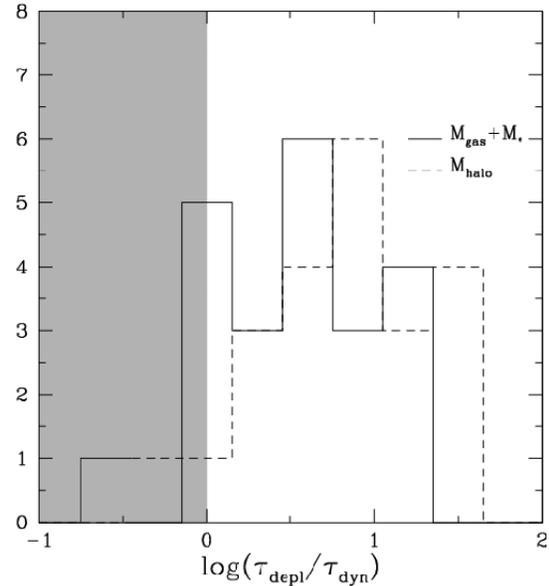}
\caption{\label{timescales} The ratio of gas depletion to dynamical timescales. The two histograms are for the case where dark matter is not included (solid line) or when it is (dashed line). The shaded region corresponds to the classical definition of a starburst. }
\end{figure} 
We finally turn to the ratio of the gas depletion timescale ($\tau_{\rm{depl}}$) to the dynamical timescale ($\tau_{\rm{dyn}}$) of a galaxy. Unlike the above two considerations, the starburst criterion is clearer here. If the current rate of SFR is such that, if sustained, the available gas reservoir will be exhausted in a time shorter than the dynamical time, then the source is bursting. The depletion timescale is simply $M_{\rm{gas}}/\rm{SFR}$. We find that, typically, $\tau_{\rm{depl}}$\,$\sim$\,8$\times$\,$10^8$\,yr. The dynamical time is defined as $\tau_{\rm{dyn}}\sim(R/GM_{\rm{tot}})^{1/2}$. To estimate this, we begin by neglecting the dark matter contribution, and calculate the dynamical time using the gas mass plus the stellar mass. We then note that including any dark matter contribution would only have the effect of decreasing $\tau_{\rm{dyn}}$ and thus any source that appears quiescent will remain so, regardless of the extra mass (we return to this point below). Using this approach we find that typically $\tau_{\rm{dyn}}$\,$\sim$\,1$\times$\,$10^8$\,yr. Fig.~\ref{timescales} shows a histogram of $\tau_{\rm{depl}}$/$\tau_{\rm{dyn}}$ using estimates both with and without dark matter in $M_{\rm{tot}}$. When the dark matter halo mass is included\footnote{Using the relation $M_{\rm{halo}}/M_{\rm{gas}}$\,$\approx$\,$35(M_{\rm{gas}}/10^7M_{\odot})^{-0.29}$ \citep{maclowfer98}}, the only starbursts in our sample are N1-013, and N1-078 (with N1-064, and N1-077 being borderline). Throughout, we assume our IRAC-based sizes, which we argued are the effective scales of the star-formation activity (see Section~5.2). The bulk of our sample, despite their being apparently somewhat more active than local, quiescent spirals, do not appear to be bona fide starbursts.
\section{Discussion}
\label{secdiscussion}

In this paper, we have presented a comprehensive study of the full infrared SEDs of a sample of 22 FIRBACK galaxies. Our new multi-component model together with the MCMC fitting technique have allowed us to test various previous assumptions about FIRBACK galaxies, as well as the spectral templates used to model them and related populations. 

We found that our galaxies have cool SEDs (low mid-IR/far-IR ratios), which remains true even for the higher luminosity sources (LIGs and ULIGs). It is not yet clear how common such sources are, and how much of a concern are they for galaxy evolution models. However, the {\sl Spitzer} 24\um\ number counts \citep{pap04} already suggest that the existing SED spectra might not be accurate in the mid-IR for the $z$\,$\sim$\,0.5\,--\,2.5 luminous and dusty sources thought to comprise the bulk of the CIB. To correct for this, the necessity for `downsizing' the mid-IR in their starburst spectra was already pointed out by \citet{lag04}. 

One of the primary questions we wanted to answer with this study was the nature of the brightest galaxies contributing to the CIB. Initially from the sub-mm data (S03) and later through optical spectroscopy follow-up (D05), it was already known that the bulk of the FIRBACK galaxies are low-$z$, moderate luminosity sources, with a small fraction of $z$\,$\sim$\,1 ULIGs . Beyond that however, little has been known of their nature, and in particular their masses, sizes, and modes of star-formation were poorly constrained. Here we have addressed these issues from a number of (admittedly not always independent) perspectives and concluded that the bulk of the sample is consistent with $\sim$\,$M^{*}$ mass galaxies, which are forming stars somewhat more actively than local spirals, but are likely not actually starbursting according to the usual definition. Their mode of star-formation is consistent with slightly enhanced activity in the disk, rather than the nuclear bursts associated with major mergers. Their cool colours also exclude any significant contribution from AGN activity. A study of the ELAIS-N2 FIRBACK galaxies by \citet{taylor05} used a set of theoretical templates for cirrus (i.e.~quiescent), starburst, and AGN galaxies. They conclude that 80\% of the FIRBACK sources are starbursts, while 20\% are cirrus galaxies. Since we find enhanced star-formation, but use more stringent starburst criteria, our conclusions are consistent. 

\subsection{Implications for the faint 24\um\ sources}
\begin{figure}
\centering
\vspace*{5.5cm}
\leavevmode
\includegraphics{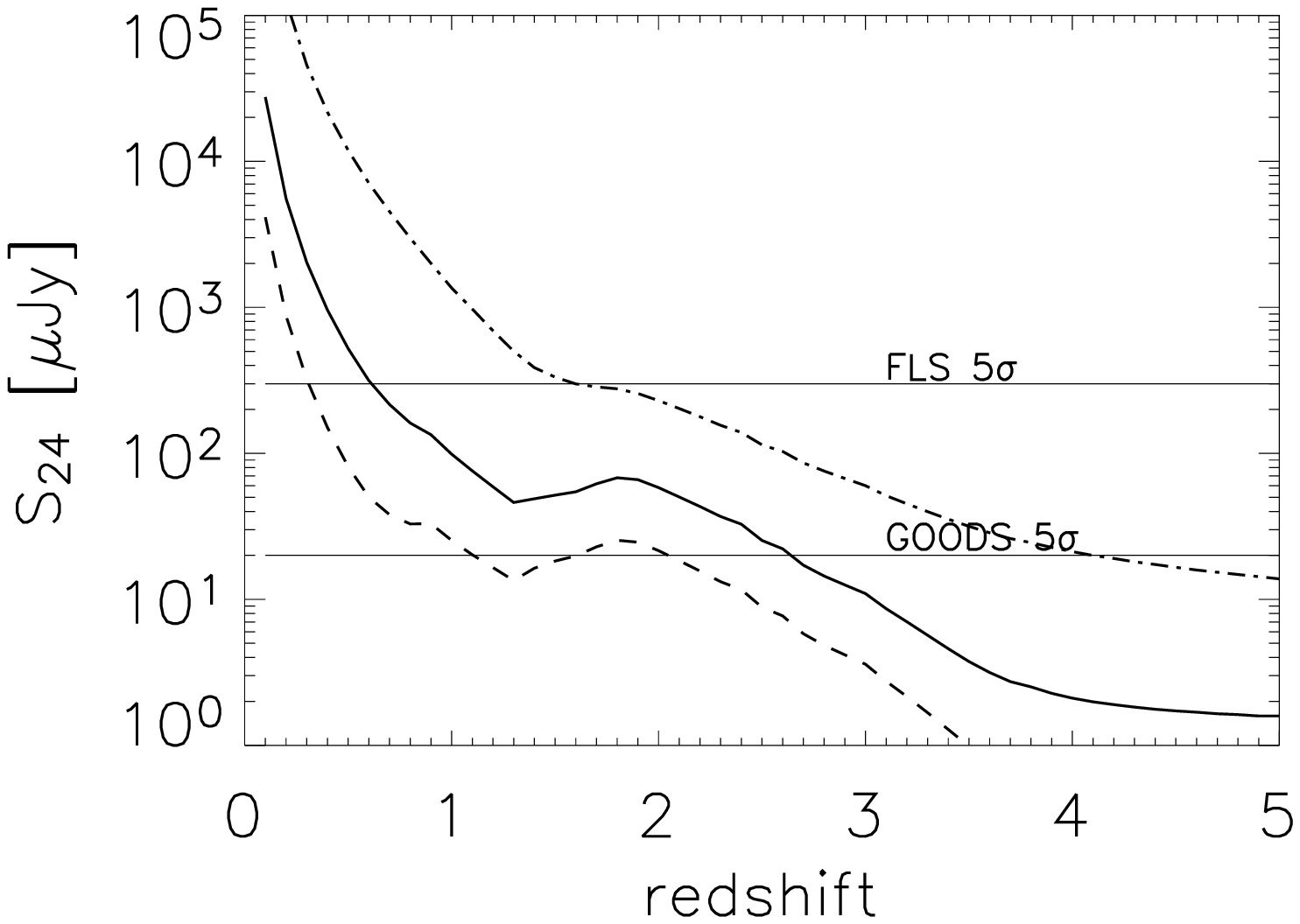}
\caption{\label{kcorr} Detectability of FIRBACK-like galaxies as a function of redshift, compared with the {\sl Spitzer} 24\um\ survey limits for SWIRE \citep{swire_cat}, and GOODS \citep{chary04}. The dashed line corresponds to the SED of a typical low-$z$ FIRBACK source (N1-007), while the solid line represents a LIG (N1-015), and the dot-dashed line is a cold ULIG (N1-064). Note that the level of PAH contribution is uncertain in the last case and therefore it might not be detectable by SWIRE at $z$\,$\sim$\,2 as suggested in this figure. }
\end{figure}

Over the past year, {\sl Spitzer} has revealed large numbers of faint (roughly $\sim$\,10--1000\,$\mu$Jy) 24\um\ sources. There has been much speculation about their nature, with obvious implications for various galaxy evolution models. In Fig.~\ref{kcorr}, we evolve the SEDs of an $L^*$ galaxy, a LIG and an ULIG from our FIRBACK sample. We compare these with the detectability thresholds of the wide and shallow SWIRE survey \citep{swire_cat} and the deepest current {\sl Spitzer} survey, GOODS-North \citep{chary04}. The SWIRE sources are expected to be predominantly low-$z$, meaning that FIRBACK-like galaxies (at only slightly higher redshifts) make up a significant fraction of them. For the deeper GOODS survey, FIRBACK analogues are expected to be a large component up to $z$\,$\sim$\,1, while in the $z$\,$\sim$\,1\,--\,2 range GOODS appears to be dominated by 11\,$<$\,$\log(L/L_{\odot})$\,$<$\,12 sources, such as the ULIG-tail of our sample, as discussed by \citet{chary04}.  Overall, Fig.~\ref{kcorr} suggests that understanding the faint 24\um\ sources requires taking into account not only M82 or Arp220 type sources, but also the much less luminous and colder FIRBACK sources, especially in the crucial $z$\,$\ls$\,1.5 regime. This is supported by the fact that while SCUBA-selected sources have close to 100\% detectability in the current generation of 24\um\ surveys \citep{egami,frayer,pope05}, most of the 24\um-selected sources do not have individual SCUBA-counterparts (stacking analyses show them to have individual 850\um\ flux $\sim$\,0.5\,mJy \citep{serjeant04}, which is below the confusion limit of SCUBA. 
\subsection{Implications for the SCUBA galaxies}
\begin{figure}
\centering
\vspace*{8cm}
\leavevmode
\includegraphics{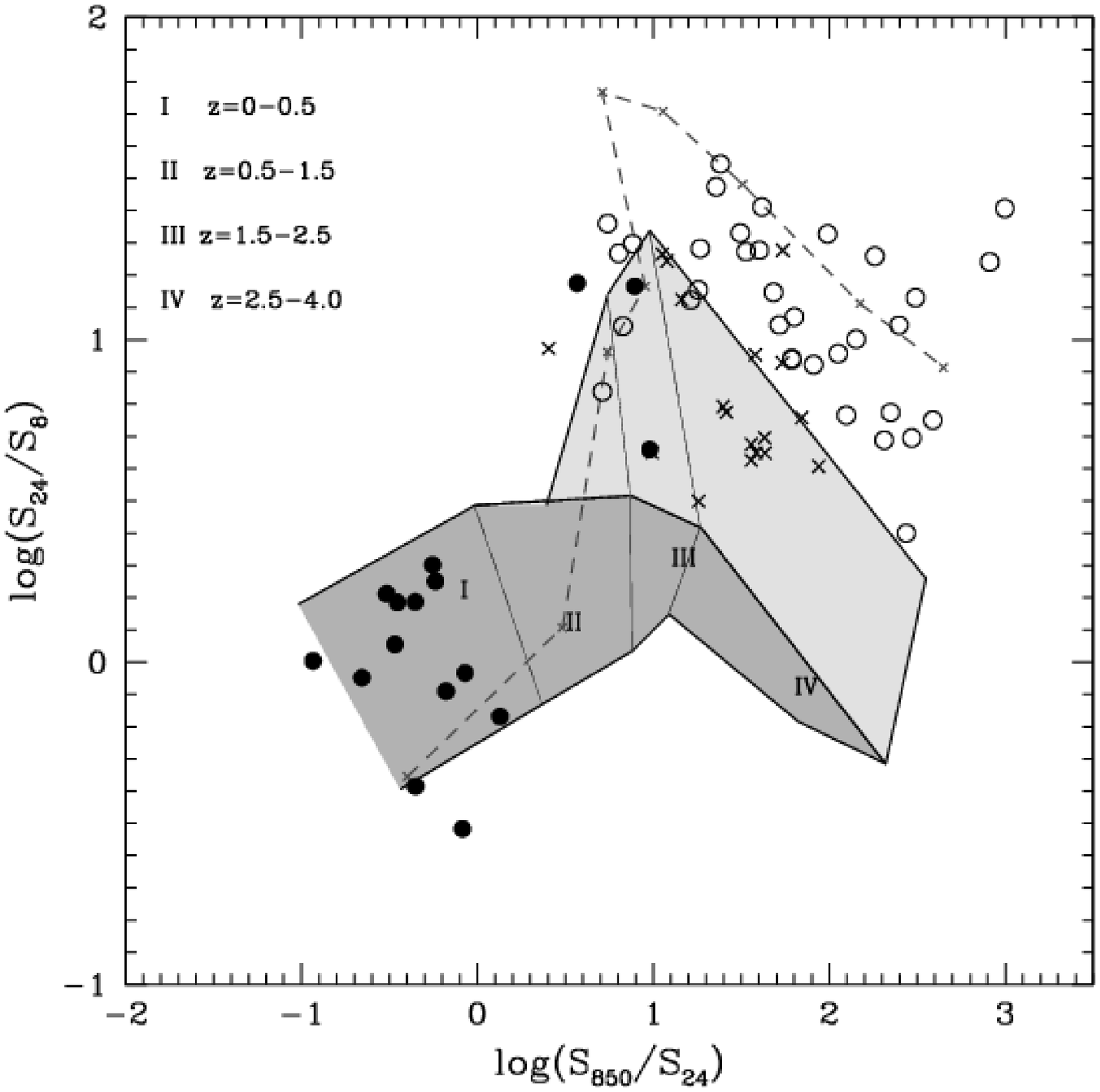}
\caption{\label{colour850_24} The evolution of two diagnostic colours: $S_{850}/S_{24}$ and $S_{24}/S_{8}$. The solid circles are our FIRBACK sample, while the open circles are SCUBA-selected galaxies from the GOODS-North field \citep{pope05}, while the crosses are other SCUBA-selected samples from shallower surveys \citep{egami,frayer}. The dark grey region represents $\log L$\,$\ls$\,11, while the light grey region shows $\log L$\,$\gs$\,11 (the boundaries should be regarded as fuzzy). The redshift evolution of the SEDs is carried out in steps of 0.5 from $z$\,=\,0 to $z$\,=\,4. The SED curves used, from top to bottom, are: N1-064, N1-029 and N1-009, and we have used them to trace the approximate boundaries for redshift evoltion, split into 4 bands as described in the legend. The dashed curve is N1-064 with an additional extinction (see Section~6.2). } 
\end{figure}

Fig.~\ref{kcorr} indicates that if the SCUBA galaxies are as cold as our ULIGs (and given that their redshifts are typically in the range $z$\,$\sim$\,2\,--\,3) then relatively shallow {\sl Spitzer} surveys (e.g. SWIRE) cannot detect any but the most luminous or the most AGN-dominated ones among them. This has implications for the {\sl Spitzer} overlap with SCUBA galaxies in wide-field sub-mm surveys, such as SHADES \citep{shades}, and those being planned with SCUBA2. The difference between the FIRBACK far-IR and blank-sky sub-mm sources is further highlighted in Fig~\ref{colour850_24}. Here we show shaded regions to indicate the colours of  normal $\log L$\,$<$\,11 galaxies, as distinct from $\log L$\,$>$\,11 ones, indicating their rough redshift evolution. The influence of the PAH feature complex as it traverses the 24\um\ filter at $z$\,$\sim$\,2 is evident. It appears that, despite some overlap with our cold ULIG template, the bulk of the SCUBA sources are redder in both colours. As the 24\um\ flux pulls them in different directions here, this rather suggests that the reddening of one colour drives the overall effect. The $S_{24}/S_{8}$ colour can be reddened by redshift, optical depth, the presence of AGN, or increased PAH strength (at $z$\,$\sim$\,2). Fully addressing the SEDs of SCUBA galaxies is clearly beyond the scope of this paper (see Pope et al. 2006). Here we merely show one scenario, i.e. that increased level of obscuration can account for the SCUBA galaxies' colours. We take our best-fit N1-064 SED ($\tau_{V}$\,=\,5), and subject it to additional extinction to a total of $\tau_{V}$\,=14. This is shown as the dashed curve in Fig.~\ref{colour850_24}. The most SCUBA galaxies now fall in-between the original and revised N1-064 template; however, we remind that we have not fully explored the parameter space, and this is merely a consistency argument.   \\   
The three FIRBACK ULIGs we have studied here are N1-040, N1-064 and N1-078. In Section~5.1, we concluded that these sources are colder, and likely have more extended star-formation activity than typically assumed for ULIGs. This is qualitatively consistent with the cirrus models of Efstathiou \& Rowan-Robinson \citep{cirrus_mrr}, which was also claimed as a possible model for the SCUBA galaxies. A small fraction of the SCUBA sources are indeed consistent with the SEDs of these galaxies, redshifted slightly to $z$\,$\sim$\,1\,--\,2. Most SCUBA galaxies, as discussed above, are consistent with the conventional view that they are highly obscured, extreme starformers (possibly including AGN) -- the result of major mergers. 

It is also worth noting that the use of the proposed 850\um\ to 24\um\ flux ratio as a redshift indicator will at best be highly unreliable, due to the 24\um\ band traversing the PAH emission and Si absorption features in the mid-IR, as well as because of the range of possible far-IR SEDs.  On the whole the SCUBA galaxies appear to exhibit a large enough range in SED shapes that applying a single model to their mid-IR photometry would be misleading. More comprehensive studies of the multi-wavelength properties of larger samples of sub-mm selected galaxies should help us understand the difference between the sources comprising the sub-mm background and those responsible for the CIB.       

\section{Conclusions}
In this paper we have combined archival {\sl Spitzer} observations of the ELAIS-N1 field with prior near-IR, far-IR, and sub-mm data in order to study the full $\sim$\,1\,--\,1000\um\ SEDs of the brightest contributors to the CIB at its peak. A novel MCMC fitting technique and a phenomenological SED model are used to make optimal use of the available data. Below we highlight the principal results of our study: \\

$\bullet$ Contrary to expectations, we have demonstrated the existence of vigorously star-forming sources, which nevertheless have low mid-IR/far-IR ratios. Our sample extends over two orders of magnitude in luminosity and includes surprisingly cold ULIGs. This is likely the result of our far-IR selection.  \\

$\bullet$ We discuss some of the issues inherent in interpreting empirical SED models. In particular, conclusions may well be influenced by the spectral sampling. Here we emphasized how limited spectral sampling and intrinsic model degeneracies might be responsible for the claimed physical \td\,--\,$\beta$ correlation.\\

$\bullet$ A number of basic parameters for our sample are derived. The typical stellar mass of our galaxies is few$\times$\,$10^{10}$\msun\ to few$\times$\,$10^{11}$\msun\ for the handful of ULIGs. Dust massses were found to be $10^7$\,--\,$10^8$\msun, where we find general consistency between emission and absorption derived values. Typical SFRs are $\sim$\,5\msun/yr. We look at combinations of the above in order to 1) check for consistency given the many inherent assumptions, and 2) allow for more meaningful comparisons with local optically-selected galaxy samples. \\

$\bullet$ In the specific SFR vs. stellar mass relation, our sample appears to have enhanced SFR with respect to local spirals, but below dwarf H{\sc ii} galaxies. The closest match to our galaxies appear to be H$\alpha$-selected galaxies. This is consistent with the emission-line spectra of our galaxies, where available (D05). \\  

$\bullet$  All our galaxies obey the Schmidt/Kennicutt relation. They appear to have enhanced star-formation activity with respect to local spirals, but less than nuclear starbursts. The more IR-luminous galaxies in our sample have higher star-formation intensity, as expected. \\

$\bullet$ We looked at the ratio of the gas depletion timescales to dynamical timescales for our sample, and find that the FIRBACK galaxies fall short of the traditional starburst definition where the gas is exhausted on timescales shorter than the dynamical time.  \\

$\bullet$ FIRBACK-like galaxies are potentially a significant component of shallow {\sl Spitzer} surveys, whose redshift distributions peak at $z$\,$\ls$\,0.3 (e.g.~FLS). Our sample does not allow us to address the Universal role of cold ULIGs, although it appears that sub-mm blank-sky surveys might also be sensitive to similarly cold (extended and/or highly obscured) sources.    

\section*{Acknowledgments}
Most of all we wish to thank the SWIRE team for obtaining, reducing and making publicly available the IRAC and MIPS data we have made use of in this project.  We are grateful to Chris Conselice for calculating the morphological parameters for our sources. We would also like to thank A. Gil de Paz for providing the data for Fig.~\ref{smass}. AS is grateful to Norm Murray and Phil Gregory for useful discussions. This research was supported by the Natural Sciences and Engineering Research Council of Canada. Lastly, we would like to thank the anonymous referee for their comments which improved the clarity of the paper.

\appendix
\section{MCMC fitting and error estimates}
\subsection{Description of our MCMC SED-fitting method}
We use a Markov Chain Monte Carlo (MCMC) \citep[see e.g.][]{gam97} approach in order to sample the posterior parameter distributions, allowing us to find the best solutions as well as the associated errors on the derived parameters. Taking {\bf a} to be the array of parameter values defining a given model, and {\bf y} to be the array of data values available, according to Bayes theorem, we have $p(\rm{\bf a}|\rm{\bf y})$\,$\propto$\,$p(\rm{\bf y}|\rm{\bf a})p(\rm{\bf a})$ where $p(\rm{\bf a}|\rm{\bf y})$ is the posterior probability distribution, $p(\rm{\bf y}|\rm{\bf a})$ is the likelihood of the data for the given model, and $p(\rm{\bf a})$ is the prior. Normalizing this (via the global probability of the data) is usually impossible in practice, since it requires knowledge of all possible models. Here, as is often the case, a single model is assumed and what we are interested in is the shape of the distribution $p(\rm{\bf a}|\rm{\bf y})$, such that the most probable values for the parameters and their errors can be derived. What we really want then, is a measure of the relative probability of a given parameter value compared with some other value. Rearranging Bayes theorem, we have:
\begin{equation}
\label{ratioeq}
\frac{p(\rm{\bf a}|\rm{\bf y})}{p(\rm{\bf a^{\prime}}|\rm{\bf y})}=\frac{p(\rm{\bf y}|\rm{\bf a})}{p(\rm{\bf y}|\rm{\bf a^{\prime}})}\frac{p(\rm{\bf a})}{p(\rm{\bf a^{\prime}})}.
\end{equation}
The ratio of priors is equal to 1 if a flat prior is assumed or is equal to $e^{a_i/a^{\prime}_i}$ if a logarithmic prior is chosen instead. Note that eq-n~\ref{ratioeq} can be used to adjust the results of a chain sampled with flat priors to test the effect of other choices of priors \citep[`importance resampling', e.g.][]{lb02}. This is the approach we take when returning to the effect of priors in Section~A.2. For now, flat priors are assumed for all parameters, and therefore from now on the posterior and likelihood distributions are used synonymously. 

In principle, arbitrary shapes of the posterior distribution can be sampled using a simple Monte Carlo approach. However for multi-dimensional problems, where the ratio of high-probability volume to total volume is very small, this can quickly become computationally prohibitive. The basic idea behind MCMC is to effectively sample this distribution by building up chains of random guesses of parameter values, where each successive guess is chosen from some much smaller proposal distribution, $q$, around the previous chain link. This move is accepted or rejected according to some criterion, which both pushes the chain toward higher probability regions, and allows for some random deviation from the straight gradient descent-type path. We follow the Metropolis-Hastings algorithm \citep{metrop53,hastings70} where all parameters are varied at once, and a guess is accepted according to the criterion: $\alpha_{i+1}$\,=\,${\rm min}[u,p_{i+1}q_{i+1}/p_{i}q_{i}]$, where the stochastic element is provided by the random number $u$\,$\epsilon$\,$[0,1]$. For the proposal distribution, we use a multivariate Gaussian, which, being symmetric, leads to $q_{i+1}/q_{i}$ of unity. Note that the exact shape of the proposal distribution is not important, but its effective width strongly affects the efficiency of the MCMC (see below). We assume that the likelihood of a given solution is the usual expression given by: 
\begin{equation}
\label{likeq}
\log \mathcal{L}=\rm{const}-\sum\left(\frac{y_{i}-y_{\rm{model}}}{\sigma_{i}}\right)^2,
\end{equation}
where the second term is the $\chi^2$. The probability ($p$) that the system finds itself in a given state is given by the Boltzman factor, where the `energy' is $\chi^2$, and thus $p_{i+1}/p_{i}$ is ${\rm exp}(-\Delta \chi^2/T)$ (this is called the `odds' of the given solution). The temperature, $T$, has effectively the same function as the width of the proposal distribution, in that it determines how easy it is for the system to jump a particular distance from its current state.  

Since brute-force MCMC is a fairly slow procedure, rather than initialize the chain at some random point we begin with some reasonable guess at the best model. Other possibilities would include simulated annealing, such as used in \citet{me05}, or equivalently using variable widths of the proposal distribution, or using any other optimization technique to find the high probability regions quickly. For our purposes here, starting with a reasonable guess is deemed sufficient since we still explore the full region of physically plausible solutions.   

The numerical parameters which need to be set are: the width of the proposal distribution for each parameter (the $q$-width); the temperature; and the overall length of the chain. Too low a $q$-width, will tend to acceptance of too many trials, while too high a value will give some jumps far outside the high probability regions, resulting in low acceptance rates. Trial-and-error has shown that an acceptance ratio in the range 10--30\% is reasonable (this is supported by empirical studies which show that $\sim$\,25\% acceptance rate in problems with $>$\,2 dimensions minimizes internal correlations in the resulting chain \citep{arate}). To find the appropriate $q$-width,  we make shorter (30,000) runs, where we vary only one parameter at a time. We start with a guess at the $q$-widths for each and incrementally adjust them until the resulting acceptance rates are 50--80\% (which is appropriate for 1-dimensional problems). By trial and error we have found that this leads to acceptance rates in the desired 10-30\% range when all seven parameters are varied simultaneously. This approach works better for our sample than fixing the widths a priori, since the dynamic range for the various parameters is too high. Such optimization of the proposal distributions before the main run speeds up the process, while preserving the ergodicity of the MCMC algorithm \citep{gs94}. We set $T$ to 0.9, which results in points $\Delta$\,$\chi^2$\,$\sim$\,1 away from the best solution to be accepted with 30\% probability. Note however, that since increasing the temperature increases the acceptance rate as well (preserving the $\Delta\chi^2/T$ ratio) and vice versa, in principle the temperature and $q$-width are degenerate, and therefore the exact value of the temperature chosen is not crucial as it will be compensated for in the above width adjustment. For the overall length of the chain, we need to find the length which both samples the posterior probability distribution well and at the same time (for practical reasons) is not much longer than what is just needed to accomplish this. We use 600,000 iterations, since our experience shows that the posterior probability around the best solution is well-sampled by this time. 

This procedure still leaves us with little sensitivity to highly disjointed solutions of equal goodness-of-fit, although this is not of concern here as such solutions will most likely converge onto unphyiscal values. 
 
\begin{figure}
\centering
\vspace*{8cm}
\leavevmode
\includegraphics{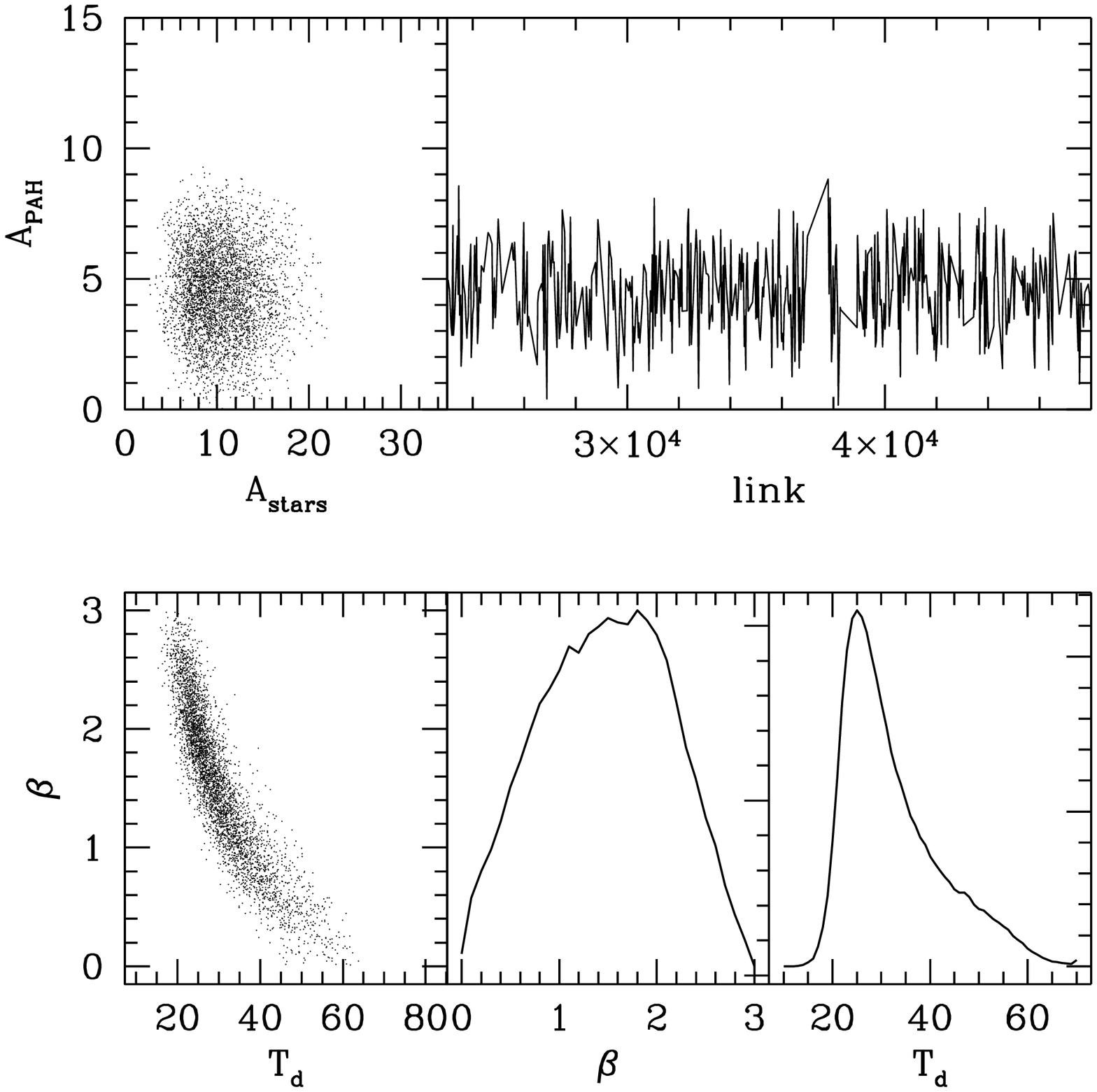}
\caption{\label{demo} Here we demonstrate what the resulting chain looks like after the procedure described. The two left-most panels show examples of both uncorrelated and correlated parameters. A thinning factor of 30 has been applied. Throughout, we plot all points with $\chi^2\leq(\chi^2_{\rm{min}}+5)$. The top right panel shows a section of the chain demonstrating the evolution of the PAH amplitude parameter. The bottom middle and right panels show the marginalized distribution for the $\beta$ and $T_{\rm{d}}$ parameters (with linear y-axes). Note the non-Gaussian shapes. }
\end{figure}

Since we start at a high-probability part of the parameter space, the `burn-in' period is less well defined than when starting at a random position, and we therefore do not formaly subtract a `burn-in' part of the chain. However, in deriving the best-fit parameters and errors below we apply a cut of $\Delta\chi^2$\,=\,5 from the minimum $\chi^2$ solution, which isolates the region of interest and thus has the same function as the `burn-in' removal. Fig.~\ref{demo} shows an example of the resulting chain in various projections.
\subsection{Error estimates}
As discussed above, the MCMC procedure samples the likelihood surface, which is proportional to the posterior probability distribution of interest. We use flat priors, where negative values are not accepted for any parameter. In addition, upper limits are set for the temperature (100\,K), optical depth (30), and $\beta$ (3). The latter is necessary, since for many sources the SNR in the sub-mm data is very low, leading to the difference in overall $\chi^2$s for unphysically large values of $\beta$ to be small, and the chain can get stuck in such regions. Since $\beta$ is believed, both on theoretical and observational grounds, to be in the range $\sim$\,1\,--\,2 \citep[see][and references therein]{dupac03}, we believe our 0\,--\,3 prior is reasonable. The temperature limit is also necessary because of the above difference in SNR. The optical depth limit is only relevant for the few sources with higher $\tau$ and faint near-IR/IRAC detections (e.g. N1-040, N1-064). These limits are imposed by returning the chain to the vicinity of its initial state in order to avoid getting stuck at the edge. We come back to the effects of the prior below. 

The chains obtained above allow for two distinct routes to obtaining the probability distributions for each parameter. The first is the easier straight marginalization of the given parameter over all the others. This is represented by: 
\begin{equation}
\label{marg}
p(a_j)da_j=\int_{i,i\neq j}^{N_{i}}p(a_i)da_i,
\end{equation}
which in practice is just counting the number of times the chain visits a particular bin of values for the given parameter $a_j$. However, since we record the $\chi^2$ value for each chain link, we can also directly obtain the likelihood distribution (see eq-n~\ref{likeq}) for the parameter. In practice, a simplified form of this is the mean likelihood distribution \citep{lb02}. The idea is to calculate the average $\chi^2$ for each bin and then compare this with the minimum $\chi^2$ achieved by the chain as
\begin{equation}
\label{meanlike}
p(a_j)\propto\exp[-(\chi^2_{j} - \chi^2_{\rm{min}})],
\end{equation}   
where $\chi^2_j$ is the mean $\chi^2$ in the $j^{\rm{th}}$ bin. 

\begin{figure}
\centering
\vspace*{9cm}
\leavevmode
\includegraphics{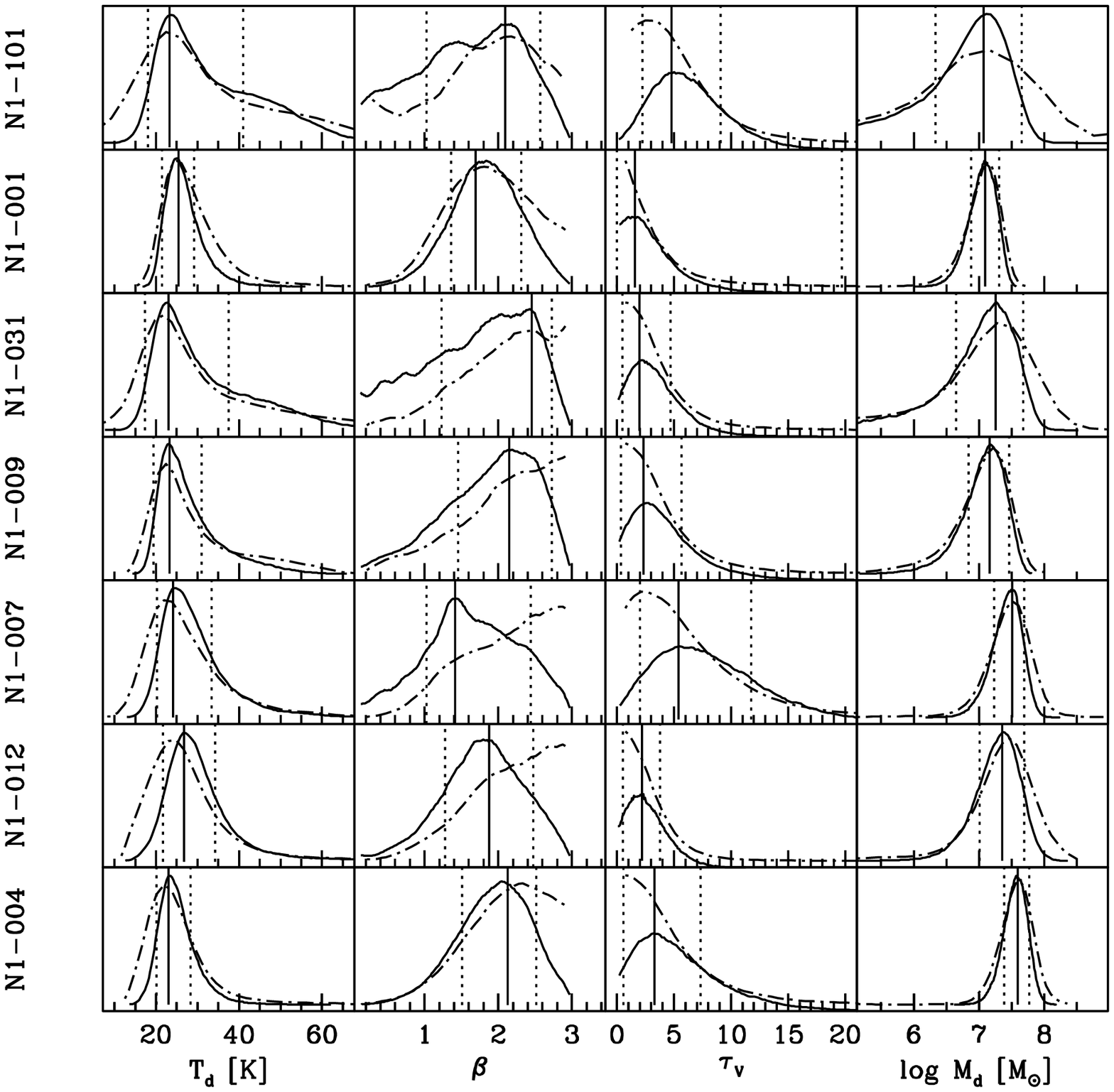}
\caption{\label{erdist} Here we compare the marginalized (solid curve) and mean likelihood (dot-dash curve) derived probability distributions for the parameters of interest. The distributions are normalized and the plots scaled to just fit the distributions as shown. The sources here and in the following figures are in order of increasing $L_{\rm{IR}}$. We use the marginalized distribution for our error estimates, where the best-fit value is given by the solid line (simply the peak of the distribution), and the 68\% confidence limits are given by the dotted lines.}
\end{figure}

In Fig.~\ref{erdist}, we show the distributions for all parameters obtained using both methods for the whole sample.  Note that since we have applied a $\chi^2$ cut (see above), secondary features in the marginalized distributions, such as blended peaks and tails, are still fairly likely. For example, in the case of N1-101, $T_{\rm{d}}$\,$\sim$\,25\,K is the preferred solution, but $T_{\rm{d}}$\,$\sim$\,40\,K is still quite likely. The uneven error-bars indicate this. Note that in Table~\ref{props}, for simplicity, we quote the average error only, but indicate the different error-bars in the figures. However, due to its definition, applying a $\chi^2$ cut when estimating the mean likelihood tends to flatten the distributions. Therefore in that case, the most `peaked' distributions possible are obtained if the whole of the chains are used. This means that parameters poorly constrained by the $\chi^2$ are clearly visible in the likelihood curves (e.g. $L_{\rm{PAH}}$ for the highest-$z$ sources N1-064, and N1-078). 
    
The best-fit values we use are merely the peaks of the above distributions. To obtain the uncertainties on those we define a desired confidence level and, starting from the peak of the distribution, move to incrementally lower equal probability bins on each side of it until the area covered is equal to the total area times the desired confidence level. Thus unsymmetric errorbars are indicative of asymmetric probability distributions.  

\begin{figure}
\centering
\vspace*{4cm}
\leavevmode
\includegraphics{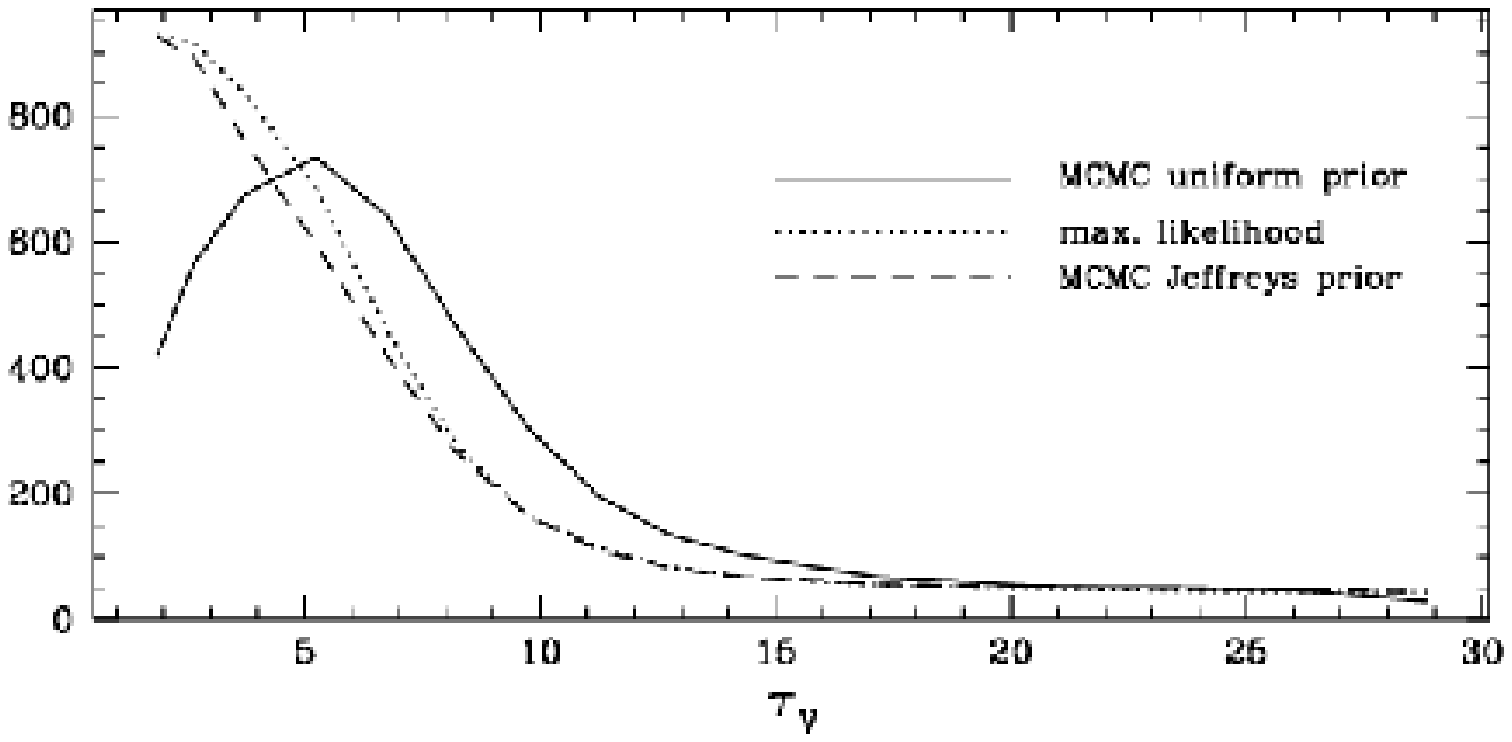}
\caption{\label{tau_prior} The effect of choice of prior on the $\tau_{V}$ probability distribution. Here we use the chain for N1-041. }
\end{figure}

In Fig.~\ref{erdist}, we see that in general the marginalized and mean likelihood distributions agree reasonably well with each other. However, there are two instances where they disagree substantially. One is for $\beta$, where the likelihood tends toward unphysically high values. The origins of this are discussed in Appendix~C. Our imposed prior confines the marginalized distribution to more reasonable values. We also find significant differences between the two approaches for $\tau_{V}$. Here we are again clearly affected by our choice of prior. Fig.~\ref{tau_prior} shows how our uniform prior-based distribution transforms into the mean likelihood distribution by application of the Jeffreys prior \citep[see e.g.][]{phil}. Note that the offsets in the PAH luminosity and stellar mass values are also due to this difference.  For the highest-$z$ sources, N1-064 and N1-078, the marginalized distributions result in essentially unconstrained $\tau_{V}$ (no clear peak in the distributions), thus the values listed in Table~2 are fairly meaningless. For these two cases, we find the mean-likelihood distribution to be more indicative, in both cases a clear, although fairly broad, peak exists at $\tau_{V}$\,$\sim$\,5.   We discuss the effects of this discrepancy as appropriate. 

\section{Photometric redshifts}
The most common approach to photometric redshift estimates is template fitting. Its effectiveness is obviously strongly dependent on the templates used and the data available \citep[see e.g.][for a discussion]{photoz_paper}. 

 Our multi-component SED model (described in Section~\ref{secmodel}) effectively represents a large and flexible template library. Therefore, we here describe the results of fitting this model, while keeping the redshift a free parameter. The results we present use the full IR SED, however, we note that the near-IR and IRAC data are the most constraining in terms of redshift determination. This is particularly true for $z$\,$\ls$\,0.3 where the major PAH features leave the IRAC 8\um\ band. With sufficiently good SNR, the approach also works at somewhat higher redshifts due to the 1.6\um\ stellar peak being probed by the near-IR data (although in our case the IRAC data is not sufficiently sensitive to do this reliably at $z$\,$\sim$\,1). We use model spectra which are generated on a logarithmic grid in wavelength, where $\Delta\log\lambda$\,=\,0.005. This leads to a minimum redshift resolution ($\Delta z/(1+z)$) of 0.01, which is adequate for our purposes.\\
\begin{figure}
\centering
\vspace*{8cm}
\leavevmode
\includegraphics{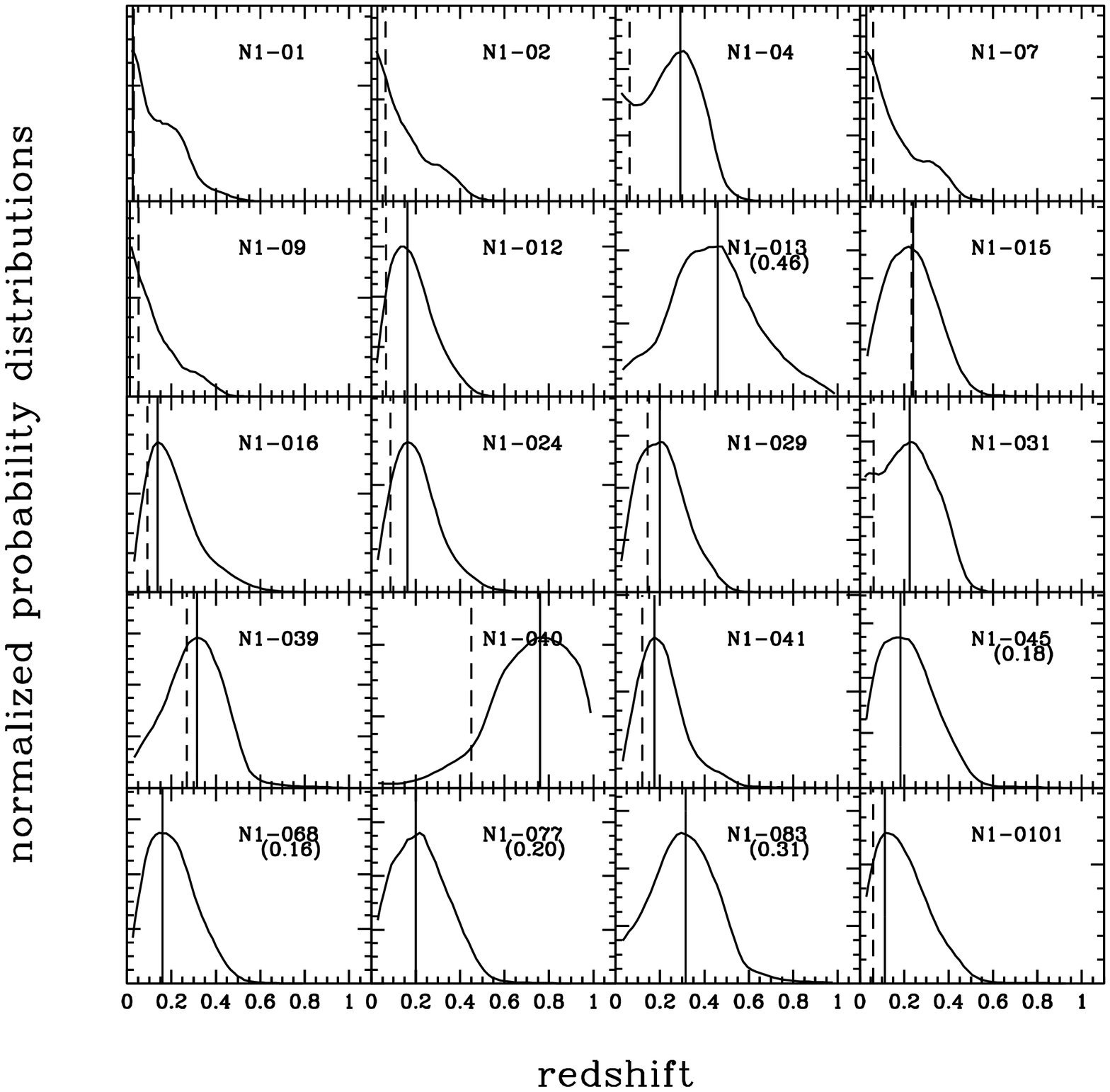}
\caption{\label{zphot_prob} The marginalized distributions for the photometric redshift estimates. The best-fit value is indicated by the solid vertical line. Where spectroscopic redshifts are available they are indicated by the dashed vertical line. Where spectroscopic redshifts are not available the best-fit photometric redshift is indicated in brackets below the name of the source. }
\end{figure} 
In order to adapt our earlier model-fitting to this problem, a few modifications are needed. First, our results in Section~4.4 suggest that $\beta$\,$\sim$\,2 adequately described the far-IR/sub-mm emission of our galaxies. Therefore, here we fix $\beta$\,=\,2. Next, to avoid some clearly unphysical solutions, we restrict the code to the range --0.1\,$<$\,$\log(A_{\rm{star}}/A_{\rm{PAH}})$\,$<$\,0.9. This condition is fully satisfied by all our sources when the spectroscopic redshifts are used (except N1-040, where PAH emission is apparently completely absent), and moreover is required if the rest-frame sources are to agree with our simulated IRAC colour-colour plot in \citet{me05}. Lastly, due to the $T/(1+z)$ degeneracy, the temperature parameter we vary here is  $T/(1+z)$. This last step removes the strongest correlation in the resulting chains. However, we find that the results are subject to additional degeneracies of redshift with other parameters, including: a negative correlation with the stellar amplitude; a positive correlation with the PAH amplitude; and a negative correlation with optical depth. These are clearly not independent. \\
Fig.~\ref{zphot_prob} shows the marginalized distributions we obtain for photometric redshift fits. Note that the mean likelihood distributions here are poorly constrained, most likely due to the above degeneracies. Overall, the agreement between our MCMC estimates and the spectroscopic redshifts (where available) is quite good. The worst case is N1-040 where the near-IR/IRAC data have very low SNR and moreover the source is apparently devoid of PAH emission, thus the constraint comes entirely from the far-IR/sub-mm which is not very accurate, as can be seen in this case (see also \citep{aretxaga}). We have left out N1-064, whose high redshift (0.91) is in a different regime (the PAH features have completely left the IRAC bands) and therefore an accurate redshift cannot be obtained with the above method. N1-078 is also believed to be at $z$\,$\sim$\,1 (S03) and has an SED very reminiscent of N1-064. Therefore we will assume it to be at $z$\,=\,0.91 as well, although note that the uncertainty is $\sim$\,0.5. 

Fig.~\ref{zphot_prob} shows that there is a bias in the model towards preferring somewhat higher redshifts compared with the spectroscopic ones. The above degeneracies result in double-peaked distributions (e.g.~N1-001, N1-007), and in some cases the `wrong' peak is preferred (N1-004, N1-031). All of these issues suggest that the model, although working in general, is imperfect at this point. Improvements are needed in minimizing the remaining degeneracies, which must involve including additional constraints. Two obvious steps are the use of the relation between absorption-derived and emission-derived dust mass, as discussed in Section~5.2 (this requires the effective sizes of galaxies), and the use of distance dimming. The latter for example effectively excludes the higher-$z$ solution for N1-004, but not for N1-031. For N1-040, the distance dimming argument would in fact support the wrong solution, which was brought about largely by the imposition of `typical' stars-to-PAH ratios. Therefore, the range in galaxy SED types needs to be known better and accounted for here before these steps can be implemented. 

After accounting for all these issues, we find that the average uncertainty of our estimated redshifts is $\langle\Delta [z/(1+z)]\rangle$\,=\,0.06. 

Thus in summary we adopt the following redshifts: N1-013(0.46), N1-045(0.18), N1-068(0.16), N1-077(0.20), and N1-083(0.31). Based on optical template fitting \citet{mrr05} derived the following redshifts: N1-013(0.58), N1-045(0.09), N1-068(0.23), N1-077(0.19), and N1-083(0.51). Assuming these are correct, our results on average are uncertain to  $\langle\Delta [z/(1+z)]\rangle$\,=\,0.07, which is slightly worse than the above quoted uncertainty as expected when comparing against other photometric redshift estimates. For simplicity, we are going to ignore the uncertainty on the estimates and treat these redshifts as fixed.

\section{The $T$\,--\,$\beta$ relation}
\label{secgrey}
A common approach, especially with a limited number of far-IR/sub-mm data points, is to fit a greybody function with characteristic temperature and emissivity. There are three main questions arising from the practice and interpretation of these fits: \\
1) what degeneracies are present in the usual formulation, and therefore what are the optimal parameters to fit?\\
2) given that this is only an approximation to the true shape of the SED, what effect does the spectral sampling (such as due to redshift) have on the results obtained? \\
3) how does one disentangle physical correlations from said degeneracies? 

For further discussion, see also \citet{blain_seds} and the alternative parameterization of \citet{baugh04}. In essence, the function one is fitting has the form: $f_{\nu}=(\nu/{\nu_0})^{\beta}$\,$\times$\,$B_{\nu}$
, where $B_{\nu}$ is the Planck function:
\begin{equation}
\label{planck}
B_{\nu}=\frac{2h}{c^2} \frac{\nu^3}{\exp(h\nu/kT)-1}.
\end{equation}

The parameters to determine here are $T_{\rm{d}}$ , $\beta$, and the nuisance parameter $\nu_0$ (or some equivalent normalization). We wish to disentangle any correlations between the three. To begin with, to understand how degeneracies arise between the various parameters, we need to consider what it is that the code is actually fitting. Intuitively, to first order this is the overall amplitude of the dust emission, the position of the peak, and the slope in the Rayleigh-Jeans tail. 

The first of these functional parameters is the bolometric dust flux, $F_{\rm{d}}$, which is given by
\begin{equation}
F_{\rm{d}}=\int f_{\nu} d\nu = \frac{2h}{c^2\nu_o^{\beta}}\left(\frac{kT}{h}\right)^{4+\beta}\Gamma(4+\beta)\zeta(4+\beta),
\end{equation}
where $\Gamma$ is the complete gamma function, and $\zeta$ is the Riemann zeta function\footnote{Note that $\zeta(4+\beta)$ is $\simeq$\,1 for any positive value of $\beta$, so does not affect the result much. The quantity $\log\Gamma(4+\beta)$ is closely approximated by 0.69(1+0.69$\beta$), which we use. }. Rearranging the above and substituting for $1/\nu_o^{\beta}$ in $f_{\nu}$ removes the bulk of the degeneracies of the normalization parameter (which is now taken to be $F_{\rm{d}}$). 

Nevertheless, the temperature and $\beta$ are still correlated. As stated above, the fitting process is also concerned with the position of the peak in the emission. By solving $df_{\nu}/d\nu$\,=\,0 we obtain:
\begin{equation}
\label{peak}
\beta=\left(\frac{h\nu_{\rm{peak}}}{kT}\right)\left[\frac{e^{h\nu_{\rm{peak}}/kT}}{e^{h\nu_{\rm{peak}}/kT}-1}\right]-3.
\end{equation}  
This equation provides a good fit to the observed correlation at higher values of $\beta$ (see Fig.~C1), but fails as $\beta$ approaches 1. The reason for this is obvious -- apart from the peak, we also need to match the Rayleigh Jeans tail. The sub-mm data do not allow $\beta$ to become arbitrarily shallow, although it can be compensated by increasing the temperature (which pulls the Planck function the other way). The constraint coming from the Rayleigh-Jeans tail can be parametrized with the sub-mm spectral index, $\alpha_{\rm{submm}}$ which is related to $\beta$, but in the Rayleigh-Jeans approximation the temperature dependence falls out leaving:
\begin{equation}
\label{alphasubmm}
\beta=\alpha_{\rm{submm}}-2\log(850/450)-3.
\end{equation}
Looking at our data (when the SNR at 450\um\ was high), we find typical values of $\alpha_{\rm{submm}}$\,$\sim$\,5, which means $\beta$\,$\sim$\,1.5 from eq-n.~\ref{alphasubmm}, which is in fact often adopted for sub-mm sources. The joint constraint on the spectral index leads to:
\begin{equation}
\label{betatot}
\beta=\left(\frac{h\nu_{\rm{peak}}}{kT}\right)\left[\frac{e^{h\nu_{\rm{peak}}/kT}}{e^{h\nu_{\rm{peak}}/kT}-1}\right]+\alpha_{\rm{submm}}-2\log\left(\frac{850}{450}\right)-6.
\end{equation} 
In Fig.~C1, we show how the above two constraints act together and therefore how the available data drive the $\beta$ and temperature values found. For example, in data without well determined peak position, assuming a low value of $\beta$ (say, 1\,--\,1.5) will automatically lead to the conclusion of hotter temperatures. Conversely, in data with cool temperatures and well determined peak, but poorly sampled sub-mm data, the conclusion will always be that $\beta$ is high (even $>$\,2 if the fit is allowed to go there). Better sampling (which could mean better SNR or more spectral points) in the sub-mm (relative to the peak) would likely change that un-physical conclusion. Conversely, as the redshift is increased, our data increasingly sample the peak rather than the tail of the thermal emission, and this has the same effect.   

\begin{figure}
\label{bscat}
\centering
\vspace*{8cm}
\leavevmode
\includegraphics{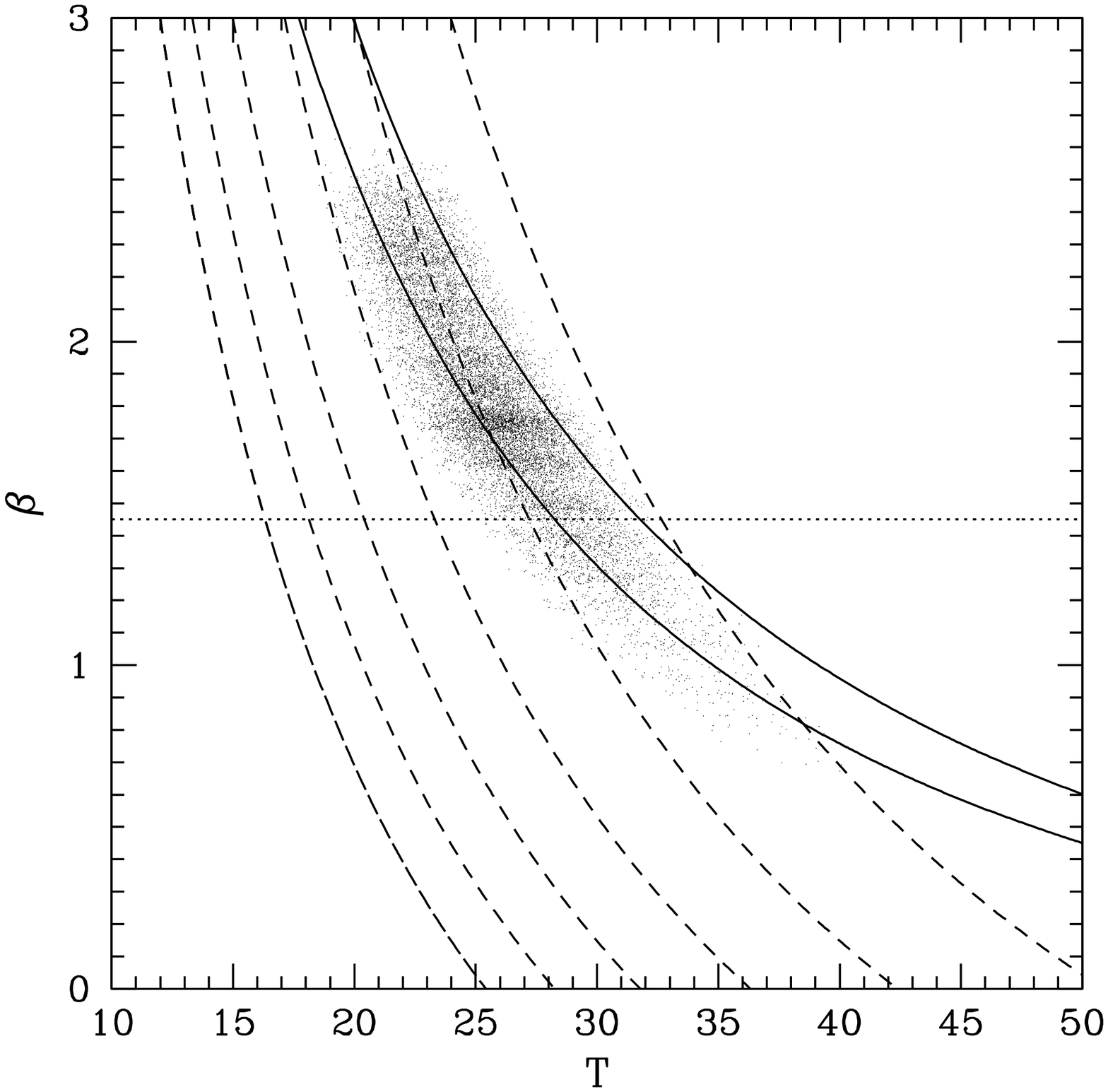}
\caption{Here we compare the relation in eq-n~\ref{betatot} and the observed degeneracy between the $\beta$ and temperature parameters. The points are from the fit to N1-024. The dashed lines show the relation in eq-n~\ref{peak}, where the location of the peak moves in increments of 20\um\ from 100\um\ to 200\um\ (top to bottom). Note the good agreement at higher $\beta$'s and the deviation at lower-valued $\beta$'s. The dotted line is the expectation of eq-n~\ref{alphasubmm} with $\alpha_{\rm{submm}}$\,=\,5 (consistent with observations). The solid lines are the combined constraint (eq-n~\ref{betatot}) for peak locations of 160\um\ and 180\um\ respectively -- the observed distribution is now reproduced much better. If not treated carefully, this parameter correlation, which exists in the fits {\it for each source}, can be misinterpreted as a physical correlation between the derived ($\beta$,$T_{\rm{d}}$) pairs {\it for a sample of sources}.}
\end{figure}

This degeneracy is empirically well known \citep{blain_seds}. Apart from this degeneracy arising from the functional form used, there has also been suggestions that a physical inverse relationship exists as well. In particular, this was seen when the best-fit temperature and $\beta$ values for different Milky Way environments were compared by \citet{dupac03}. The relative homogeneity of our sample (see Fig.~\ref{betat}) does not allow us to firmly support or reject this relationship (although the apparently warmest source, N1-078, supports it). A wider range of galaxy types with well sampled SEDs would be needed to address this firmly. However, we note that (as discussed above) as the peak of the SED shifts, the combined $\beta$,\,$T$ distribution moves as well. But in the Dupac et al.~sample the shortest wavelength is 100\um, thus whenever the peak is shortward of that (corresponding to $\sim$\,30\,K for $\beta$\,$\sim$\,2), the sub-mm index begins to dominates the fit, flattening the distribution. At the other extreme for example, as the peak reaches the coldest observed values (corresponding to peaks at $\sim$\,200\um), we are sampling a distribution where $T$\,$\sim$\,15\,K for $\beta$\,$\sim$\,2 (see Fig.~C1). Therefore, it appears that this correlation is easily explained as a sequence of progressively colder/hotter environments. The apparent $\beta$\,--\,$T$ correlation is merely a combination of the intrinsic correlation between $T_{\rm{d}}$ and $\beta$ for individual sources and limited wavelength coverage. That being said, a physical correlation here is to be expected from the different optical properties of grains at different temperatures. But it is unclear that the single greybody approach is capable of testing this adequately.

\bsp

\label{lastpage}

\end{document}